\documentclass[superscriptaddress,nofootinbib,11pt,
      twoside,preprintnumbers,notitlepage]{revtex4}
\usepackage{epsfig,amsmath,amssymb}
\usepackage{color}
\usepackage{float}
\usepackage[colorlinks]{hyperref}
\usepackage{subfigure}
\usepackage[utf8]{inputenc}
\usepackage{slashed}
\usepackage{times}

\hypersetup{
    linktocpage,
     colorlinks,
     citecolor=darkgreen,
     linkcolor= darkgreen,
     urlcolor=darkgreen
}
\definecolor{darkred}{rgb}{0.5,0.0,0.0}
\definecolor{darkblue}{rgb}{0.0,0.0,0.9}
\definecolor{darkerblue}{rgb}{0.0,0.0,0.5}
\definecolor{purple}{rgb}{0.5,0.0,0.5}
\definecolor{darkgreen}{rgb}{0.0,0.5,0.0}
\definecolor{black}{rgb}{0.0,0.0,0.0}
\definecolor{brown}{rgb}{0.6,0.4,0.2}
\definecolor{newpurple}{rgb}{0.65, 0.38, 0.61}
\definecolor{newyellow}{rgb}{0.9718, 0.6093, 0.0759}
\definecolor{amber}{rgb}{1.0, 0.75, 0.0}
\definecolor{newblue}{rgb}{0.4, 0.52, 0.85}
\definecolor{newred}{rgb}{0.8524, 0.2595, 0.3294}
\definecolor{newgreen}{rgb}{0.2, 0.8, 0.2}
\definecolor{SMgreen}{rgb}{0.56, 0.69, 0.19}
\definecolor{neworange}{rgb}{0.94, 0.462, 0.162}

\definecolor{BrickRed}{rgb}{0.9,0.1,0}

\newcommand{\bea}{\begin{eqnarray}}
\newcommand{\eea}{\end{eqnarray}}
\newcommand{\beq}{\begin{equation}}
\newcommand{\eeq}{\end{equation}}
\newcommand{\ec}{\end{center}}
\newcommand{\bc}{\begin{center}}
\newcommand{\gev}{\ensuremath{\,}{\rm GeV}}
\newcommand{\mev}{\ensuremath{\,}{\rm MeV}}

\newcommand{\fig}[1]{Fig.~\ref{#1}}
\newcommand{\eq}[1]{Eq.~(\ref{#1})}
\newcommand{\eqsand}[2]{Eqs.~(\ref{#1}) and (\ref{#2})}

\begin{document}

\preprint{TTP-20-002}
\title{Higgs decay into a lepton pair and a photon revisited}
\author{Aliaksei Kachanovich, Ulrich Nierste, and Ivan Ni\v sand\v
  zi\'c} 

\email[Electronic addresses:]{aliaksei.kachanovich@kit.edu,
  ulrich.nierste@kit.edu, ivan.nisandzic@kit.edu}

\affiliation{Institut f\"ur Theoretische Teilchenphysik (TTP),
  Karlsruher Institut f\"ur Technologie (KIT), 76131 Karlsruhe, Germany}


\begin{abstract}
  We present new calculations of the differential decay rates for
    $H\to \ell^+\ell^- \gamma$ with $\ell=e$ or $\mu$ in the Standard
    Model. The branching fractions and  forward-backward asymmetries, defined
    in terms of the flight direction of the photon relative to the
    lepton momenta, depend on the cuts on energies and invariant masses
    of the final state particles.  
    For typical choices of these cuts we find the branching ratios
   $B(H\to e \bar e \gamma)=5.8\cdot 10^{-5}$ and $B(H\to \mu \bar \mu
    \gamma)=6.4\cdot 10^{-5}$ and the forward-backward asymmetries
   $\mathcal{A}^{(e)}_{\text{FB}}=0.343$ and $\mathcal{A}^{(\mu)}_{\text{FB}}=0.255$. We provide compact analytic
    expressions for the differential decay rates for the use in
    experimental analyses. 
\end{abstract}

\maketitle

\section{Introduction}
Since the discovery of a Higgs boson with a mass of 125$\,$\gev\
\cite{Chatrchyan:2012xdj,Aad:2012tfa} in 2012, the LHC experiments CMS
and ATLAS put a major effort into the precise determination of its
couplings.  The Standard Model (SM) accommodates a minimal Higgs sector,
with just one Higgs doublet, and it is natural to ask whether nature
foresees a richer Higgs sector than the SM. A possible imprint of an
extended Higgs sector are deviations of the measured couplings from
their SM predictions \cite{Englert:2014uua}. To date CMS and ATLAS have
studied the couplings of the discovered Higgs boson to $W$
\cite{Aad:2015ona,ATLAS:2014aga,Sirunyan:2018egh,Sirunyan:2019twz} and
$Z$
\cite{Chatrchyan:2013mxa,Sirunyan:2019twz,Aad:2014tca,Aaboud:2018ezd}
bosons, $\tau$ leptons
\cite{Aad:2015vsa,Aaboud:2018pen,Sirunyan:2017khh}, $b$
\cite{Sirunyan:2018kst,Aaboud:2018zhk} and $t$
\cite{Sirunyan:2018shy,Aaboud:2018urx} quarks, as well as photons
\cite{Sirunyan:2018ouh,Aaboud:2018xdt}.  The latter coupling is
loop suppressed and has been probed though the rare decay
$H\to \gamma\gamma$.  Rare Higgs decays are especially sensitive to
physics beyond the SM and even probe scenarios with only one Higgs
doublet as in the SM. For instance, $H\to \gamma\gamma$ data were
instrumental to rule out a fourth sequential fermion generation from a
global analysis of Higgs signal strengths \cite{Eberhardt:2012gv}.
Phenomenological analyses of two-Higgs-doublet models (2HDM) usually
assume simple versions of the Yukawa sector (called type I,II,X, or Y),
in which different observables become correlated and the largest
imprints are on the heavy fermions of the third generation
\cite{Eberhardt:2013uba, Belanger:2013xza, Baglio:2014nea,
  Kanemura:2014bqa, Dev:2014yca, Broggio:2014mna, Chowdhury:2015yja,
  Bernon:2015qea, Haber:2015pua, Han:2017pfo, Chowdhury:2017aav}. In
such models the dynamics of light fermions of the first and second
generation follow the pattern of the third generation and measurements
of Higgs decay rates involving light fermions will only provide
redundant information. However, as outlined in the following paragraph
there are well-motivated phenomenological reasons to consider the
possibility that physics beyond the SM shows imprints on the decays of
the 125$\,$\gev\ Higgs bosons into final states containing light fermions. It
is therefore mandatory to measure the corresponding decay rates
accurately and to compare the data with precise SM predictions.

In this paper we study the SM predictions for the rare decays
$H\to \ell^+\ell^-\gamma$ with focus on $\ell=e$ and $\ell=\mu$. While
the amplitude of $H\to \ell^+\ell^-$ is suppressed by one power of the
Yukawa coupling $y_\ell=m_\ell/v$, where $m_\ell$ is the lepton mass and
$v=174\gev$ is the vacuum expectation value (vev) of the Higgs field,
the radiative analogue does not suffer from this suppression.

Electroweak loop contributions instead involve the Higgs coupling to
heavy gauge bosons or to the top quark and permit a nonzero decay
amplitude even for $y_\ell=0$, producing the lepton pair in a state with
angular momentum $j=1$. The decay rate $\Gamma (H\to e^+e^-\gamma)$
exceeds $\Gamma (H\to e^+e^-)$ by far, while
$\Gamma (H\to \mu^+\mu^-\gamma)$ and $\Gamma (H\to \mu^+\mu^-)$ are
comparable in size. In the case of $\Gamma (H\to \tau^+\tau^-\gamma)$
the electroweak loop contribution is much smaller than the tree-level
contribution proportional to $y_\tau^2$, which simply amounts to the
bremsstrahlung contribution to $H\to \tau^+\tau^-$. It is important to
note that $H\to \ell^+\ell^-$ and $H\to \ell^+\ell^-\gamma$ probe
different sectors of beyond the Standard Model (BSM) models (chirality-flipping
vs\ chirality-conserving couplings to lepton fields) and are therefore
complementary. A further motivation to study
$H\to \mu^+\mu^-\gamma$ is the $3.7\sigma$ discrepancy between the
measured anomalous magnetic moment of the muon, $a_\mu$, and its SM
prediction \cite{Keshavarzi:2018mgv}. $a_\mu$ involves the magnetic
operator $\bar L_\mu \Phi \sigma_{\alpha\beta}\mu_R \, F^{\alpha\beta}$,
with the lepton doublet $L_\mu=(\nu_\mu ,\mu)$, the Higgs doublet
$\Phi$, and the electromagnetic field strength tensor
$F^{\alpha\beta}$. BSM models with loop contributions to the coefficient
of the magnetic operator may as well affect the
$H$-$\bar\mu$-$\mu$-$\gamma$ couplings. 
Another related topic are the hints of 
violation of lepton flavour universality encoded in the ratios 
$R_{K^{(*)}}\equiv B(B\to K^{(*)} \mu^+\mu^-)/B(B\to K^{(*)} e^+e^-)$
\cite{Aaij:2017vbb,Aaij:2019wad}, which support BSM physics coupling 
to left-chiral leptons \cite{Alguero:2019ptt,Aebischer:2019mlg}.
Also here the underlying BSM dynamics can eventually be tested with 
$H\to \ell^+\ell^-\gamma$.
None of the $H\to \ell^+\ell^- \gamma$ 
decays has been observed yet;  cf.~Ref.~\cite{Sirunyan:2018tbk} for LHC limits.

Analytic expressions for differential $H\to \ell^+\ell^-\gamma$ decay
rates have been derived in Refs.~\cite{Abbasabadi:1996ze}
\cite{Chen:2012ju}. 
After the discovery of the 125\,\gev\
Higgs boson Ref.~\cite{Abbasabadi:1996ze} was updated
\cite{Dicus:2013ycd} and two new detailed analyses based on novel
calculations of the decay rate have been presented in
Refs.~\cite{Passarino:2013nka} \cite{Han:2017yhy}. Comparing the decay rates
$d\Gamma (H\to \ell^+\ell^-\gamma)/d \sqrt s$, where $\sqrt s$ is the
invariant mass of the lepton pair, presented in these papers, we find
significant discrepancies, which motivates the new calculation of this
decay rate presented in this paper.  We use a linear $R_\xi$ gauge, so
that we can use the vanishing of the $W$ and $Z$ gauge parameters as
a check of our calculation. This check is especially valuable in the
context of the $Z$ width, which must be taken into account when the
invariant mass of the lepton pair is close to the $Z$ boson mass and
special care is needed to ensure a gauge-independent result
\cite{Denner:1999gp,Denner:2006ic}. In this kinematic region our decay
is indistinguishable from $H\to Z\gamma$ (for recent LHC search limits
cf.~\cite{Aaboud:2017uhw,Sirunyan:2017hsb}). In
Ref.~\cite{Passarino:2013nka} it is argued that $H\to Z\gamma$ is not a
properly defined physical process and one should instead discuss the
full decay chain, including the $Z$ decay, such as
$H\to \ell^+\ell^-\gamma$.  Since we keep the gauge parameters
arbitrary, we can track how the unphysical gauge-dependent pieces of the
$H\to Z[\to \ell^+\ell^-]\gamma$ subprocesse cancel with those of other diagrams.
We will quote a compact formula for the differential decay
rate $d^2\Gamma (H\to \ell^+\ell^-\gamma)/(ds\, dt)$ with respect to the
Mandelstam variables $s$ and $t$, where $\sqrt t$ is the invariant mass
of the $(\ell^-,\gamma)$ pair, and discuss the forward-backward
asymmetry of the photon.

Our paper is organized as follows: In the following section we present 
our calculation and discuss our results, including a comparison with the
literature. Sec.~\ref{conc} contains our conclusions, followed by two
appendices guiding through our analytic results.

\section{Calculation and results}

\subsection{Amplitudes\label{amp}}
\begin{figure}[t]
\hrule
	\begin{center}
		\subfigure[t][]{\includegraphics[width=0.22\textwidth]{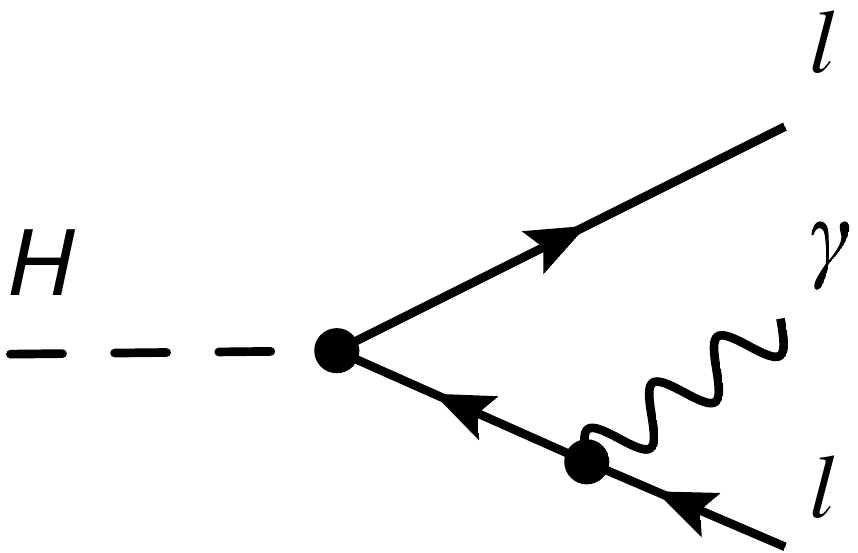}}
		\hspace{.6cm}
		\subfigure[t][]{\includegraphics[width=0.22\textwidth]{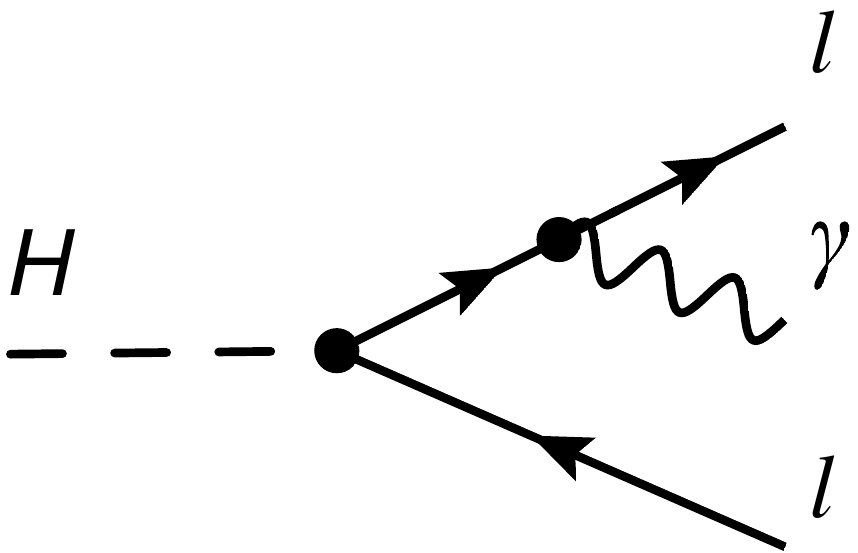}}
         \end{center}
	\caption{Tree-level Feynman diagrams.
	}
	\label{Fig:1}
~\\[-3mm]\hrule
\end{figure}

The amplitude for the tree-level photon emission process (see Fig. \ref{Fig:1}) is
\begin{eqnarray}
  \mathcal{A}_{\rm tree} = -\frac{e^{2} m_{\ell} \varepsilon_{\nu}^\ast
  (k)}{2 m_{W} 
  \sin\theta_{W}}\bigg[\frac{\overline{u}(p_{1})(\gamma^\nu \slashed{k} 
  +2\,p_{1}^\nu)v(p_2)}{t-m_\ell^2}-\frac{\overline{u}(p_1)(\slashed{k}
  \gamma^\nu 
  + 2\,p_2^\nu) v(p_2)}{u-m_\ell^2}\bigg]\, , \label{tree}
\end{eqnarray}
with our conventions for the kinematical variables as following: We
denote four-momenta of photon, lepton and antilepton by $k$, $p_1$, $p_2$,
respectively. Squared invariant masses are denoted by the Mandelstam
variables $s=(p_1+p_2)^2$, $t=(p_1+k)^2$, and $u=(p_2+k)^2$ obeying the
relation $s+t+u=m_H^2+2m_\ell^2$, where $m_H$ is the Higgs boson mass.
$e$, $m_W$, and $\theta_W$ are the electromagnetic coupling constant,
mass of the $W$ boson, and weak mixing angle, respectively. $u$ and $v$
are the lepton and antilepton spinors and $\varepsilon$ is the
polarization vector of the photon. 

\begin{figure}[bt]
\hrule\medskip
	\begin{center}
		\subfigure[t][]{\includegraphics[width=0.2\textwidth]{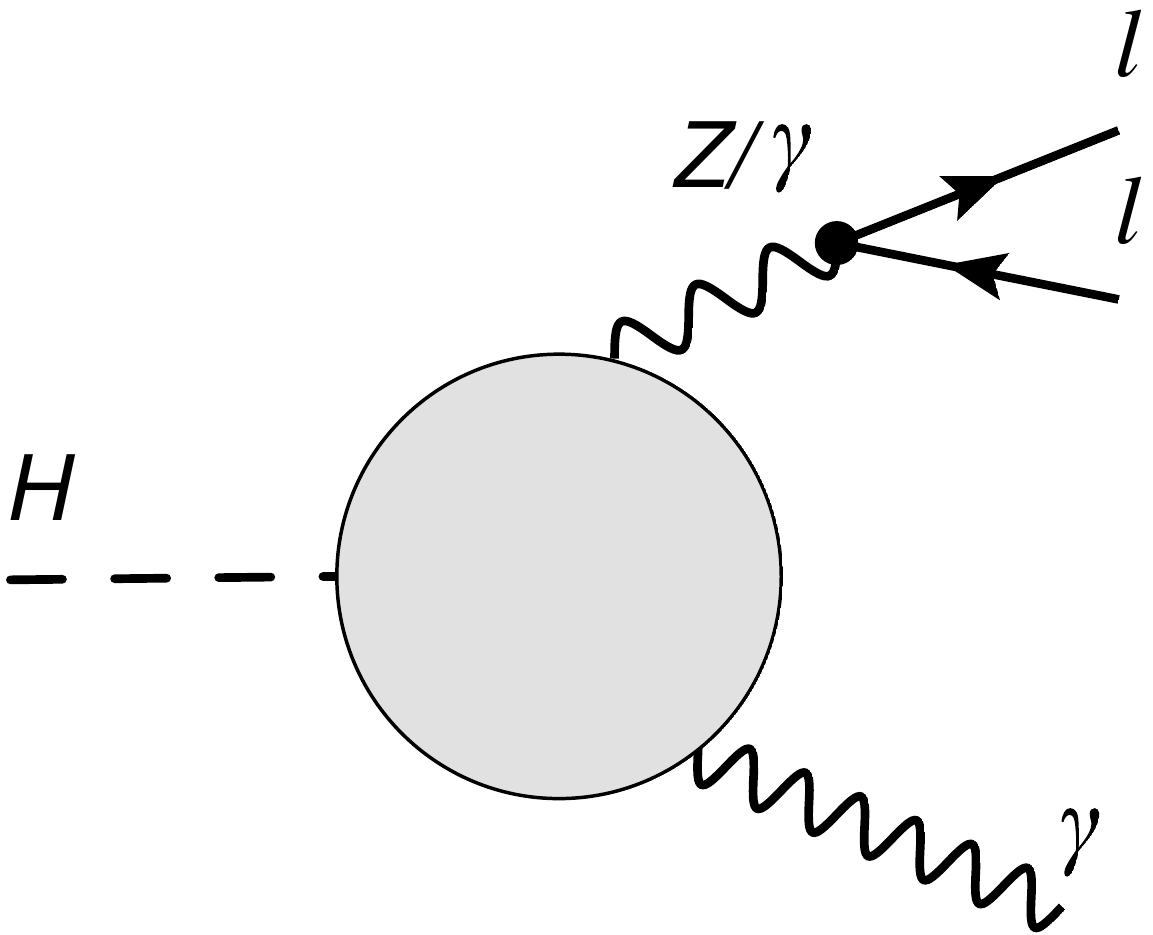}}
		\hspace{.6cm}
		\subfigure[t][]{\includegraphics[width=0.2\textwidth]{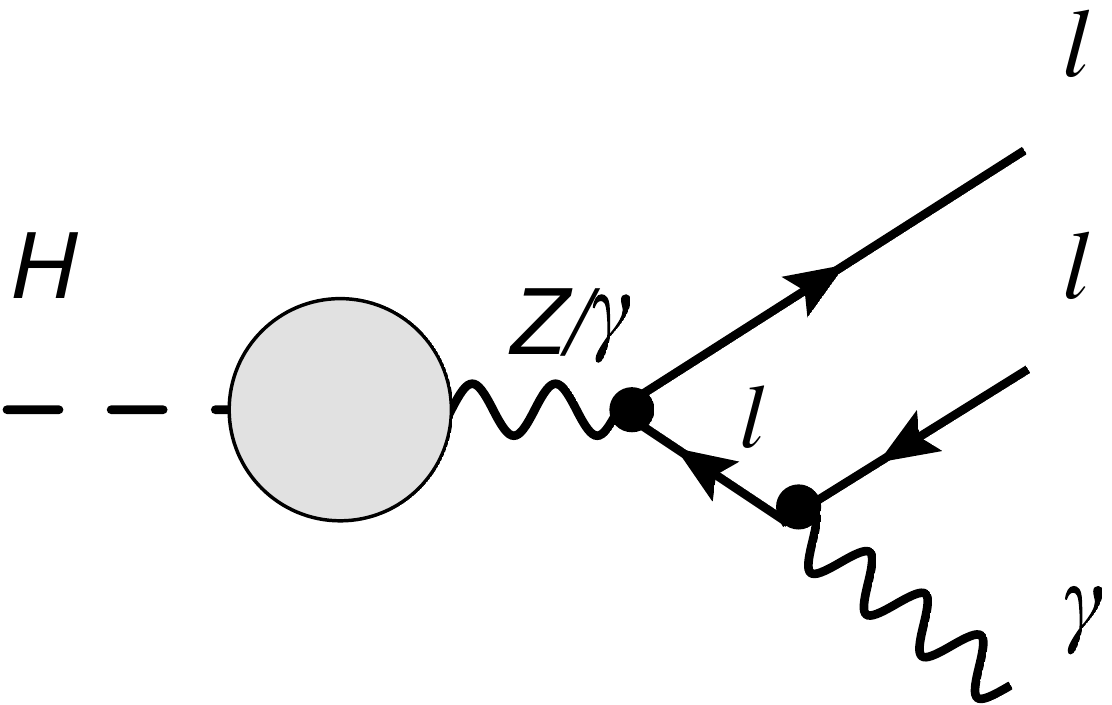}}\\
		\hspace{.6cm}
		\subfigure[t][]{\includegraphics[width=0.2\textwidth]{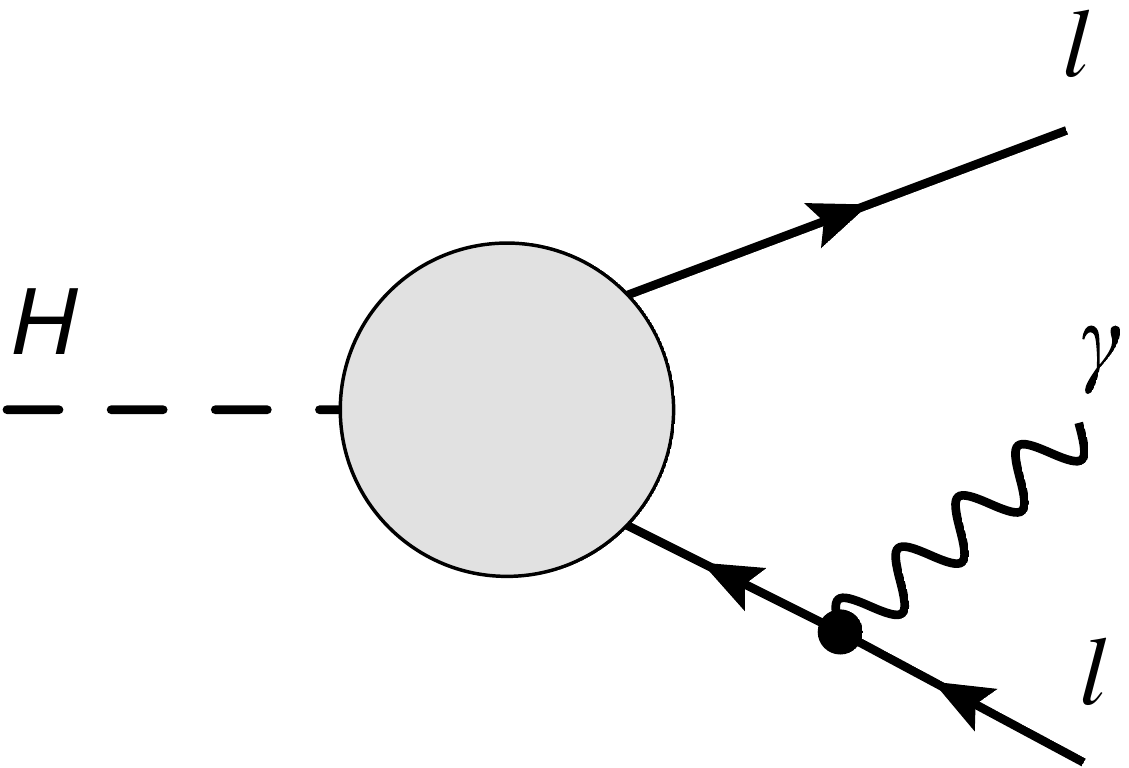}}
		\hspace{.6cm}
		\subfigure[t][]{\includegraphics[width=0.2\textwidth]{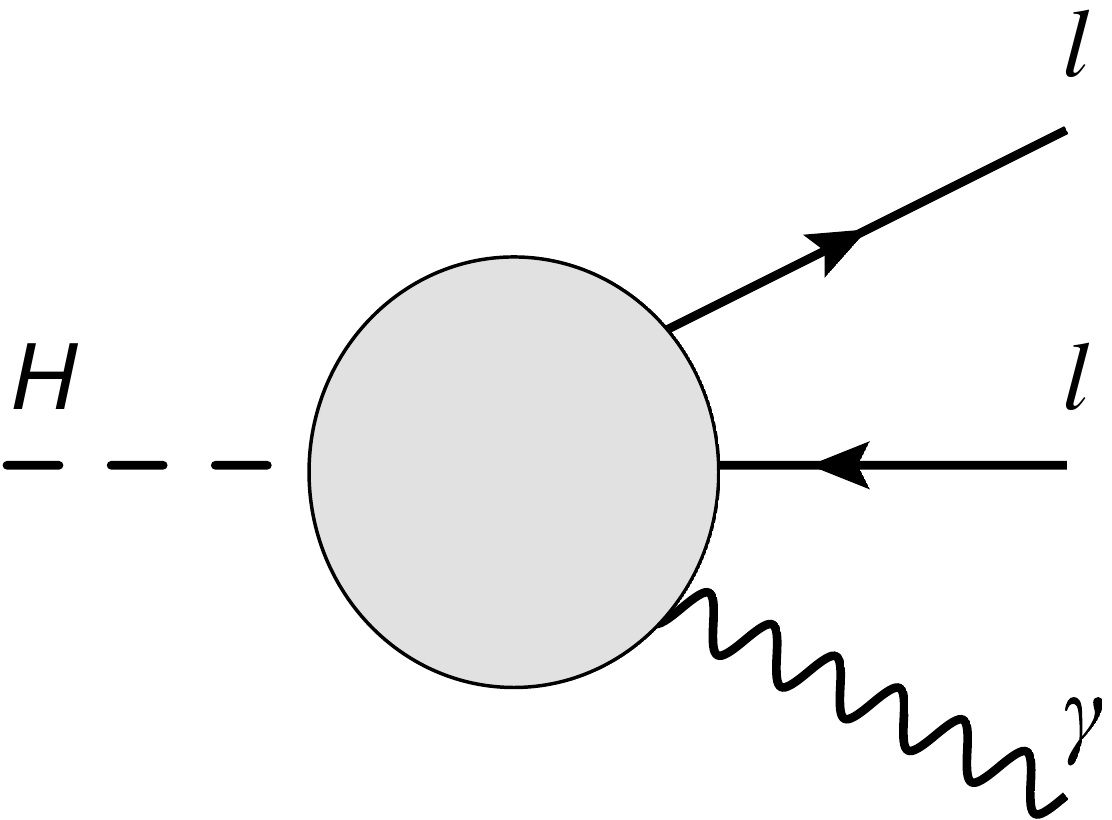}}
         \end{center}
         \caption{Schematic representation of the classes of one-loop
           diagrams that contribute to $H\to \ell\ell\gamma$ at the one-loop
           level, with the blob representing one-particle irreducible
           diagrams. Total contribution from class (b) is vanishing.
           None of the classes shown in (a), (c), and (d) is separately gauge independent within linear $R_\xi$ gauge.  
             }
	\label{Fig:generic}
~\\[-3mm]\hrule
\end{figure}

\begin{figure}[tb]
\hrule
	\begin{center}
		\subfigure[t][]{\includegraphics[width=0.22\textwidth]{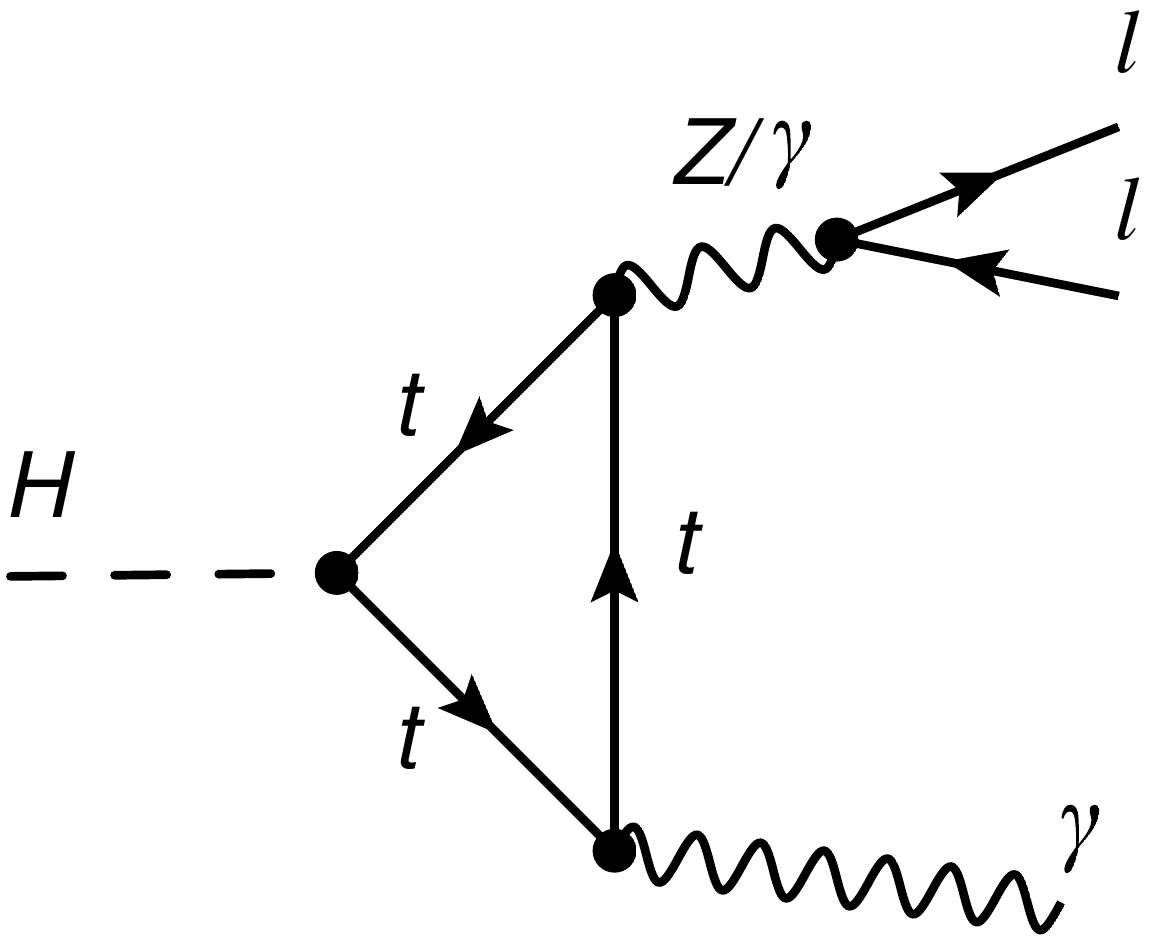}}
		\hspace{.6cm}
		\subfigure[t][]{\includegraphics[width=0.22\textwidth]{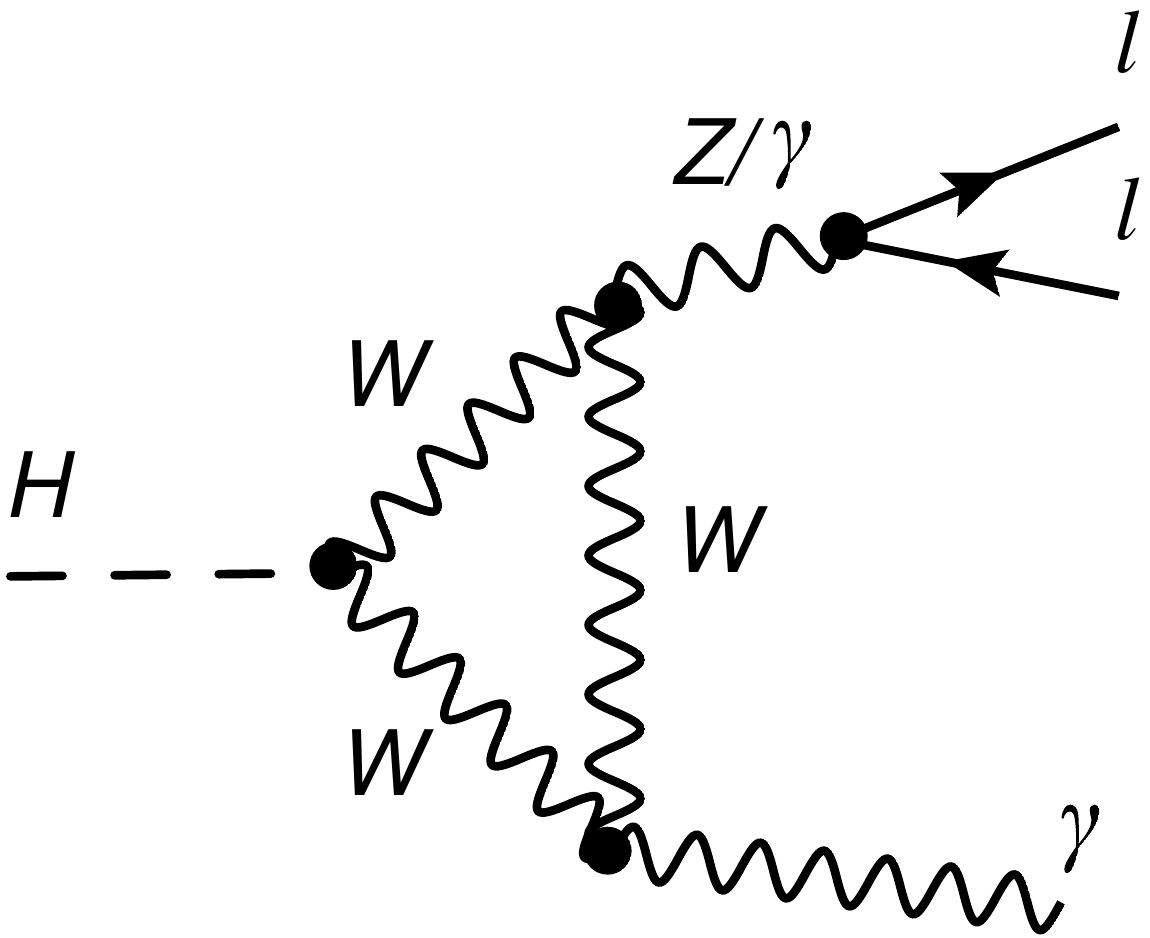}}
		\hspace{.6cm}
		\subfigure[t][]{\includegraphics[width=0.22\textwidth]{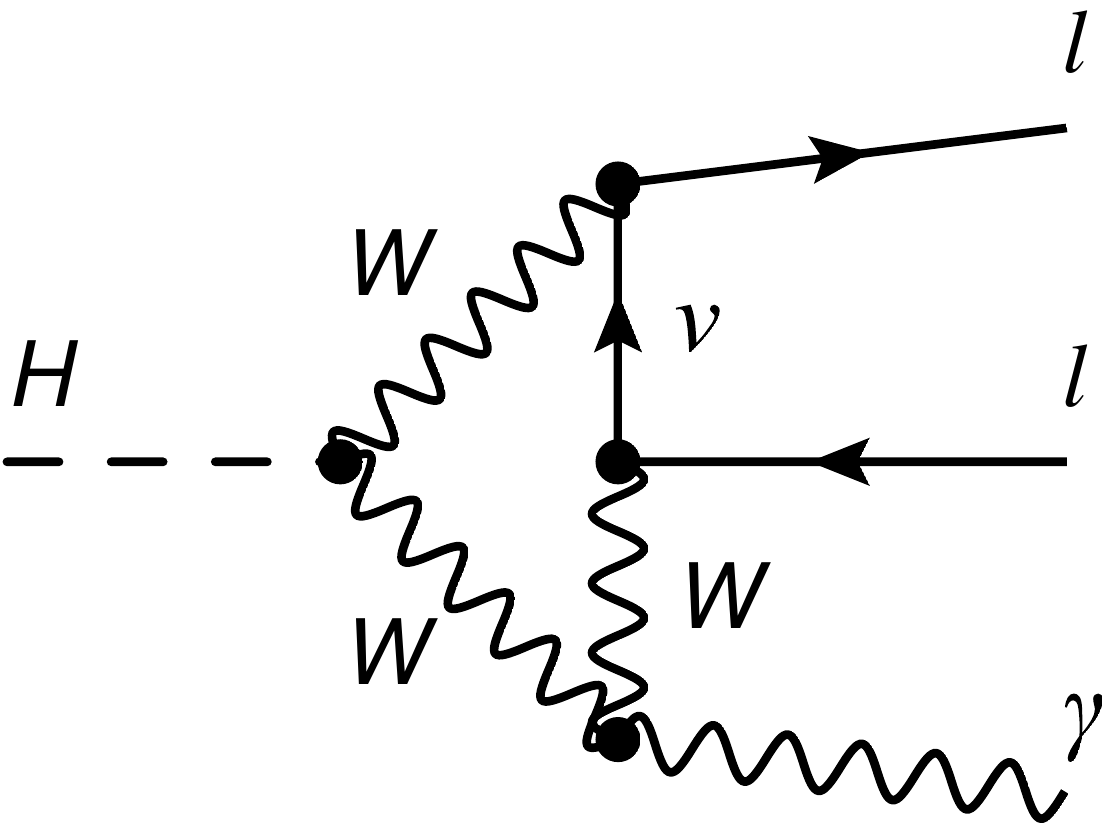}}\\
		\hspace{.6cm}
		\subfigure[t][]{\includegraphics[width=0.22\textwidth]{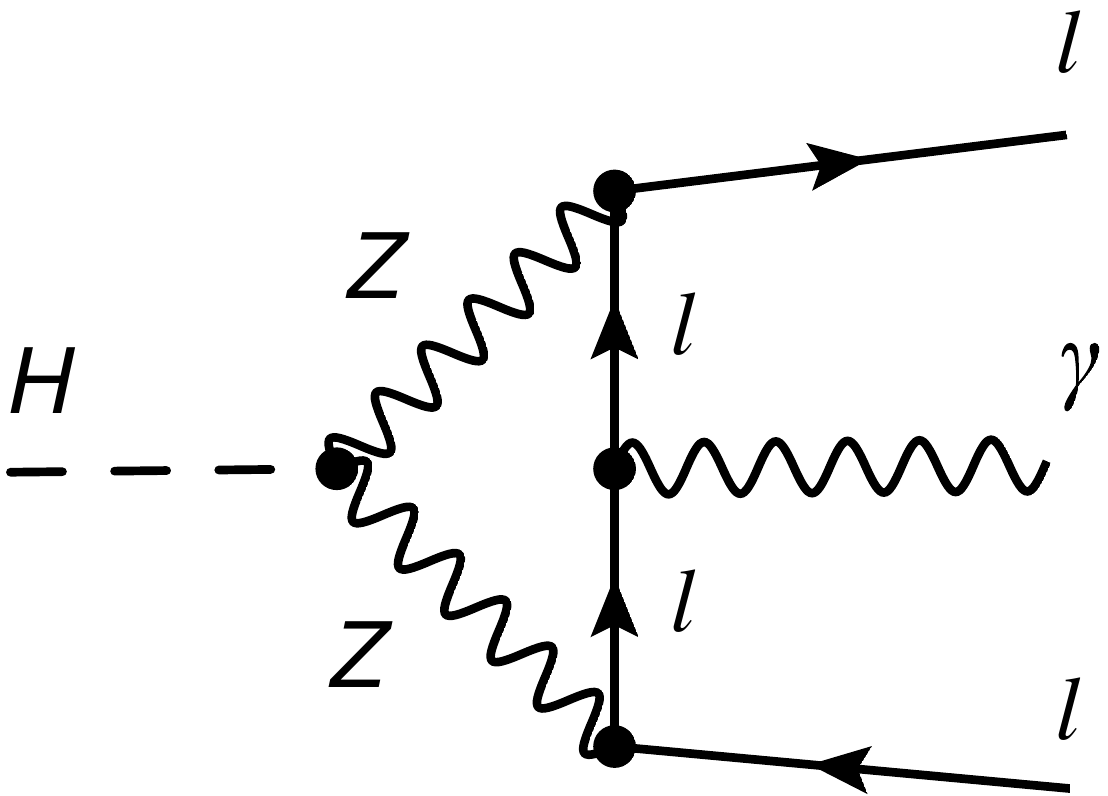}}
		\hspace{.6cm}
		\subfigure[t][]{\includegraphics[width=0.22\textwidth]{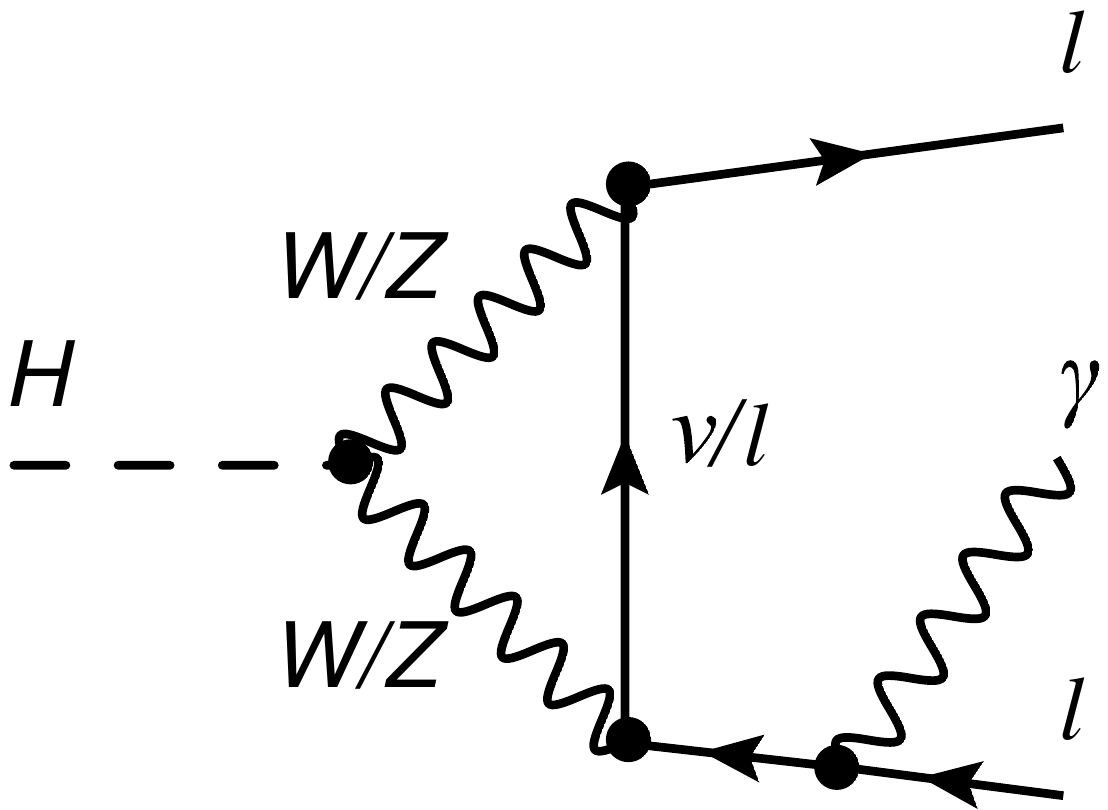}}
		\hspace{.6cm}
		\subfigure[t][]{\includegraphics[width=0.2\textwidth]{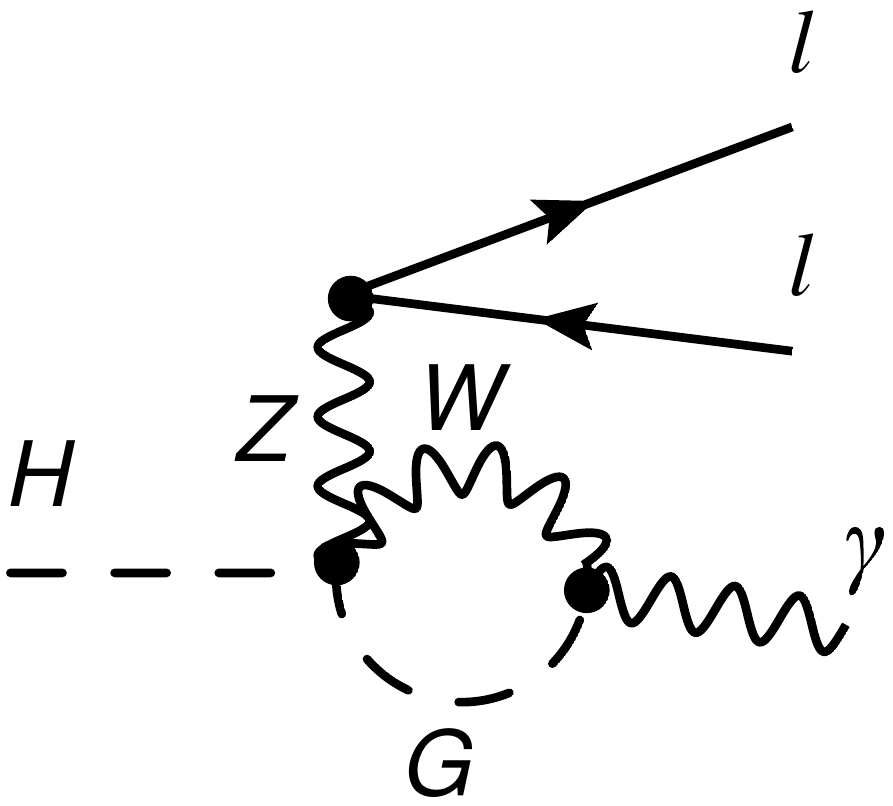}}
         \end{center}
	\caption{Sample Feynman diagrams contributing to $H\to \ell\ell\gamma$ at one-loop level.
	}
	\label{Fig:2}
~\\[-3mm]\hrule
\end{figure}

In the one-loop contribution we can neglect $y_\ell$. The Feynman
diagrams may be grouped into several classes as depicted in \fig{Fig:generic}.
Sample diagrams can be found in \fig{Fig:2}.
The one-loop amplitude can be parametrized as 
\begin{eqnarray}
  \mathcal{A}_{\text{loop}}&=&\big[(k_\mu\,p_{1\nu}-g_{\mu\nu}\,
                                k\cdot p_1)  \bar{u}(p_2)\big(a_1
                                \gamma^\mu P_R 
   + b_1 \gamma^\mu P_L\big) v(p_1)\nonumber\\
                            &+&(k_\mu\,p_{2\nu}-g_{\mu\nu}\, k\cdot p_2)
                                \bar{u}(p_2)
 \big(a_2\gamma^\mu P_R + b_2\gamma^\mu P_L\big)v(p_1)\big]
 \varepsilon^{\nu\,\ast}(k)\,.
\label{loop_amp}
\end{eqnarray}
$P_{L,R}=(1\mp\gamma_5)/2$ are chiral projectors and the coefficients
$a_{1,2}$ and $b_{1,2}$ are functions of $s,t,u$ and the particle
masses.  We present the analytic results for $a_{1}$ and $b_{1}$ in
Appendix~\ref{sec:res} in \eqsand{a1}{b1}; $a_2$ and $b_2$ are obtained by
interchanging $t$ and $u$: $a_2(t,u)=a_1(u,t)$ and
$b_2(t,u)=b_1(u,t)$. The compact results in \eqsand{a1}{b1} involve the
coefficient functions appearing in the Passarino-Veltman decomposition
\cite{Brown:1952eu,Passarino:1978jh} of the tensor integrals.  A result
fully reduced to scalar one-loop functions
\cite{tHooft:1978jhc,Denner:1991qq} is given in the ancillary file
attached to this paper in terms of a \emph{Mathematica} file.

We keep only the top quark in the fermion triangle diagrams while
neglecting all other  Yukawa couplings.  This leaves us with $119$
one-loop diagrams. We have explicitly checked that the final result 
is finite in the soft and collinear limits.  This infrared (IR) safety
must hold, because for massless leptons the one-loop result constitutes
the leading contribution and there are no diagrams with virtual photon
to cancel any IR divergences.

We now comment on several differences with respect to existing results
in the literature. Refs.~\cite{Chen:2012ju,Sun:2013rqa} contain additional
terms of the form
$\epsilon_{\mu\nu\rho\sigma}k^\rho (p_1+p_2)^\sigma
\varepsilon^{\nu\,\ast}$ involving the Levi-Civita tensor in the final
result for the loop amplitude.  This contribution is absent in our
result in Eq. \eqref{loop_amp}\footnote{These terms are also absent 
in the final results of Refs.~\cite{Abbasabadi:1995rc,
    Abbasabadi:1996ze}}. We have found that this term indeed appears in
the top-quark triangle diagram (a) of Fig.~\ref{Fig:2}, but cancels with
the corresponding diagram with opposite fermion number flow. 
We have used the 't Hooft-Veltman scheme for the treatment of $\gamma_5$ in
$D$-dimensions. 

Using a nonlinear gauge \cite{Abbasabadi:1995rc} the authors of 
Ref.~\cite{Abbasabadi:1996ze} have identified classes of diagrams
which separately satisfy the electromagnetic Ward identity. 
Using instead the usual linear $R_\xi$-gauge we find a straightforward cancellation of
the $Z$ boson gauge parameter $\xi_Z$, while the   cancellation of
the dependence on the $W$  boson gauge parameter $\xi_W$ 
involves subtleties:
The tree-level relation $e^2/g_2^2=\sin^2\theta_W=1-m_W^2/m_Z^2$ (which is
promoted to an all-order relation in the on-shell renormalization scheme
for $\sin\theta_W$) is instrumental for the cancellation of $\xi_W$ from the
result. But in order to describe the decay distribution for
$\sqrt{s}\equiv m_{\ell\ell}$ in the region around the $Z$ resonance, we
must use a Breit-Wigner shape for the $Z$-propagator in the diagrams of
class (a) in \fig{Fig:generic}. Yet this modification spoils the
cancellation of $\xi_W$ dependence between triangle and other diagrams; a
remedy is the use of the complex-mass scheme introduced in
\cite{Denner:2006ic,Denner:1999gp} as e.g.\ done in
Ref.~\cite{Passarino:2013nka}. We instead start with strictly real gauge
boson masses, verify the $\xi_W$ independence of the result, and
subsequently add the finite $Z$ width $\Gamma_Z$ to the final,
gauge-independent result. 

{In most phase space regions our loop functions are real; exceptions
  are kinematical situations such as $t>M_W^2$ or $u>M_W^2$ permitting
  on-shell cuts of the loops. Switching to the complex mass scheme makes
  the real loop expressions develop imaginary parts proportional to the
  gauge boson width; in the corresponding phase space regions the decay
  rates found in the two approaches differ by terms quadratic in
  $\Gamma_{W,Z}$. {We checked that the difference in
    $d\Gamma (H\to \ell^+\ell^-\gamma)/d \sqrt s$ between the two
    approaches is numerically negligible.}}  Therefore the treatment of
$\Gamma_Z$ cannot be the reason for the numerical differences between
our result and the various results in the literature.

We next shortly describe  the tools used in our calculation. We
have generated the Feynman diagrams with the \emph{FeynArts}\ package
\cite{Hahn:2000kx}. For the evaluation of the loop integrals and the
reduction to scalar basis functions we have used \emph{FeynCalc}\
\cite{Shtabovenko:2016sxi, Mertig:1990an}\ and have verified the
cancellation of infrared poles in the final result using Package-X
\cite{Patel:2015tea} linked to \emph{FeynCalc}\ via the
\emph{FeynHelpers}\ package \cite{Shtabovenko:2016whf}.  Then we have
evaluated the loop functions using \emph{CollierLink}\
\cite{Patel:2015tea, CollierLink} that provides a \emph{Mathematica} link to
the \emph{Collier}\ package \cite{Denner:2016kdg, Denner:2005nn,
  Denner:2010tr}. The \emph{LoopTools}\ package \cite{Hahn:1998yk, LoopManual} is
finally used for additional numerical checks. For the numerical
integration over the phase space variables we use \emph{Vegas}\
\cite{Lepage:1977sw} from the \emph{Cuba}\ library \cite{Hahn:2004fe},
and \emph{Mathematica}\ \cite{Mathematica}.

\subsection{Decay rates}
\begin{figure}[tb]
\hrule
	\begin{center}
		\subfigure[t][]{\includegraphics[width=0.488\textwidth]{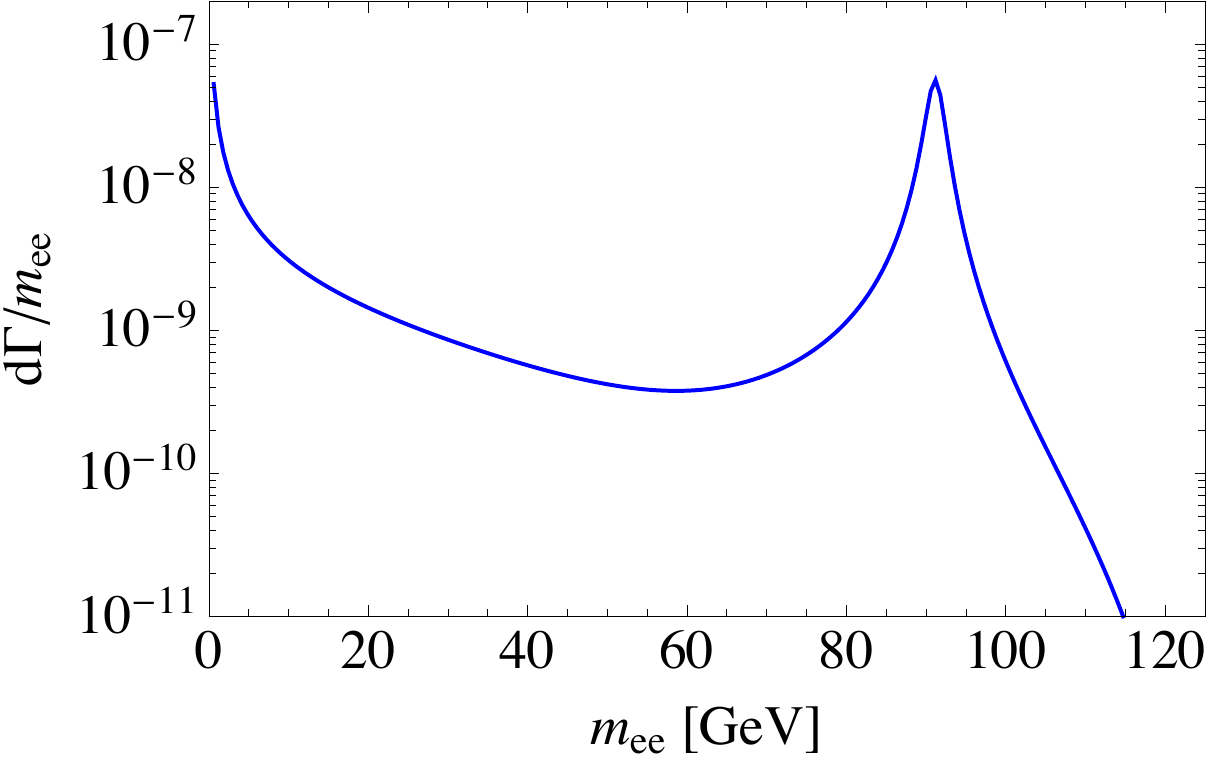}}
		\hspace{.1cm}
		\subfigure[t][]{\includegraphics[width=0.488\textwidth]{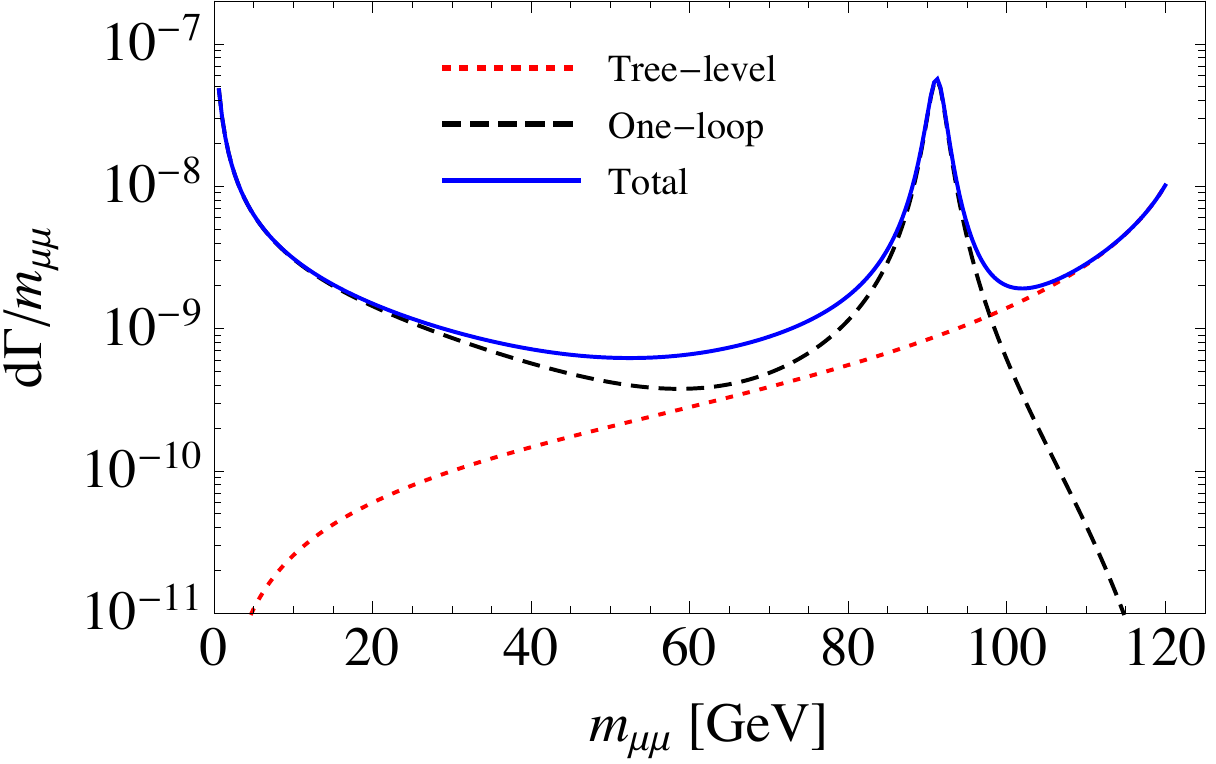}}
         \end{center}
         \caption{Differential decay rate with respect to the invariant dilepton
           mass for (a) electrons and (b) muons. The tree-level,
           one-loop, and total contributions are denoted by red dotted,
           black dashed lines, and solid blue lines, respectively. The tree-level
           contribution for the case of electrons is negligible. The
           only cut which we impose for these plots is
           $E_{\gamma\text{,\,min}}=\,5\,\text{GeV}$ which merely
           amounts to a lowering of the maximum value of 
           $m_{\ell\ell}$.  }
	\label{Fig:3}
~\\[-3mm]\hrule
\end{figure}
The tree contribution to the decay rate 
shown in \fig{Fig:1} reads 
\begin{eqnarray}
\displaystyle \frac{d^2\Gamma_{\text{tree}}}{ds\,dt} &=& \mathcal{N}\Big[\frac{9 m_\ell^4+m_\ell^2(-2s +t -3 u)+ t~u}{(t-m_\ell^2)^2}+\frac{9 m_\ell^4+m_\ell^2(-2s +u -3 t)+ t~u}{(u-m_\ell^2)^2}\nonumber \\
&+&\frac{34 m_\ell^4-2 m_\ell^2(8s+5(t+u))+2(s+t)(s+u)}{(t-m_\ell^2)(u-m_\ell^2)}\Big]\, ,\label{tree-decayrate}
\end{eqnarray}
with
\begin{equation}
\label{norm}
\mathcal{N} =  \frac{e^{4} m_{\ell}^{2}}{256\,\pi^{3} \sin^{2}\theta_{W} m_{W}^{2} m_{H}^{3}}\,.
\end{equation}
The contribution from the one-loop diagrams is
\begin{equation}
\frac{d^2\Gamma_{\text{loop}}}{ds\,dt}=\frac{s}{512 \pi^3 m_H^3}\big[t^2(\vert a_1\vert^2 + \vert b_1\vert^2)+u^2 (\vert a_2\vert^2+\vert b_2\vert^2)\big]\, .
\label{loop-decayrate}
\end{equation}
While we set $m_\ell$ to zero in \eq{loop-decayrate}, we retain a
nonvanishing value of $m_\ell$ in the kinematical limits of the
phase-space intergration.  The limits for the variables $s$ and $t$ can
be expressed as 
\begin{eqnarray}
s_{\text{min}}=4\,m_\ell^2, \quad s_{\text{max}} = m_H^2, \qquad
\displaystyle  t_{\text{min}(\text{max})}&=\frac{1}{2}\bigg(m_H^2 - s + 2m_\ell^2\mp (m_H^2 - s)\sqrt{1 - 4 m_\ell^2/s}\bigg).
\end{eqnarray}

The tree-level contribution exhibits an infrared pole as 
$s$ approaches its maximum value $s_{\text{max}}=m_H^2$ that corresponds
to a vanishing photon energy $E_\gamma$ in the Higgs boson rest-frame,
$E_\gamma= (m_H^2-s)/2m_H$. For the evaluation of the total decay rate
we impose the cut $E_{\gamma\text{,\,min}}$ which lowers the maximum
value of $s$ to $s_{\text{cut}}=m_H^2-2m_H E_{\gamma\text{,\,min}}$. The
resulting differential decay rate over $m_{\ell \ell}=\sqrt{s}$, for
$\ell=e,\mu$, is shown in \fig{Fig:3}. One notes the enhancement
from the $Z$-pole as well as the tail of photon pole starting at
$m_{\ell\ell,\,min}$. The contribution of the interference between
tree-level and one-loop contributions is negligible, as well as
the effect of the tree-level diagrams  in the case of electrons.
For the evaluation of the full decay rates we employ the kinematical
cuts of Refs.~\cite{Dicus:2013ycd,Passarino:2013nka}, namely: 
\begin{align}
& s, t, u > (0.1\,m_H)^2,\qquad 
  E_\gamma > 5\,\text{GeV},\\\nonumber
&  (E_{1} > 7 \,\text{GeV},\qquad 
  E_{2} > 25\,\text{GeV})\quad\text{or}\quad (E_{1} > 25 \,\text{GeV},\quad 
  E_{2} > 7\,\text{GeV}).   
\end{align}

We use the following input for the physical parameters:  
\begin{equation}
\begin{split}
  &\qquad  m_W = 80.379\,
  \text{GeV}\,,\qquad m_Z = 91.1876\,\text{GeV}\,,\qquad \sin^2\theta_W = 1-\frac{m_W^2}{m_Z^2}=0.223013\,,\\
&  \qquad\qquad 
  m_t = 173.1\,\text{GeV} \,,\qquad m_H = 125.1\,\text{GeV}\,, \\
  & G_F=1.1663787\times 10^{-5}\,\text{GeV}^{-2}\,,\qquad \alpha^{-1} =
  \frac{\pi}{\sqrt{2}G_F m_W^2 \sin^2\theta_W} = 132.184\,.
\label{inp}
\end{split}
\end{equation}
Here $m_t$ is the top mass, $G_F$ is the Fermi constant, $\alpha=e^2/(4\pi)$ is the fine
structure constant and the other quantities are
defined after \eq{tree}. Our calculation of  $\alpha$ in \eq{inp}
  employs the tree-level relations between $G_F$ and the fundamental
  parameters of the SM, as e.g.\ in Ref.~\cite{deFlorian:2016spz}. Radiative corrections shift this value to
  the familiar $\alpha^{-1}\simeq 128$, but the ambiguity stemming from
  the choices of numerical values for  $\alpha$ and other inputs can only be
  resolved by performing a next-to-leading-order (NLO) (e.g.\ two-loop) calculation of the
   $H\to \ell^+\ell^-\gamma $ decay rate.

\begin{figure}[tb]
\hrule ~\\[2mm]
	\begin{center}
		\subfigure{\includegraphics[width=0.52\textwidth]{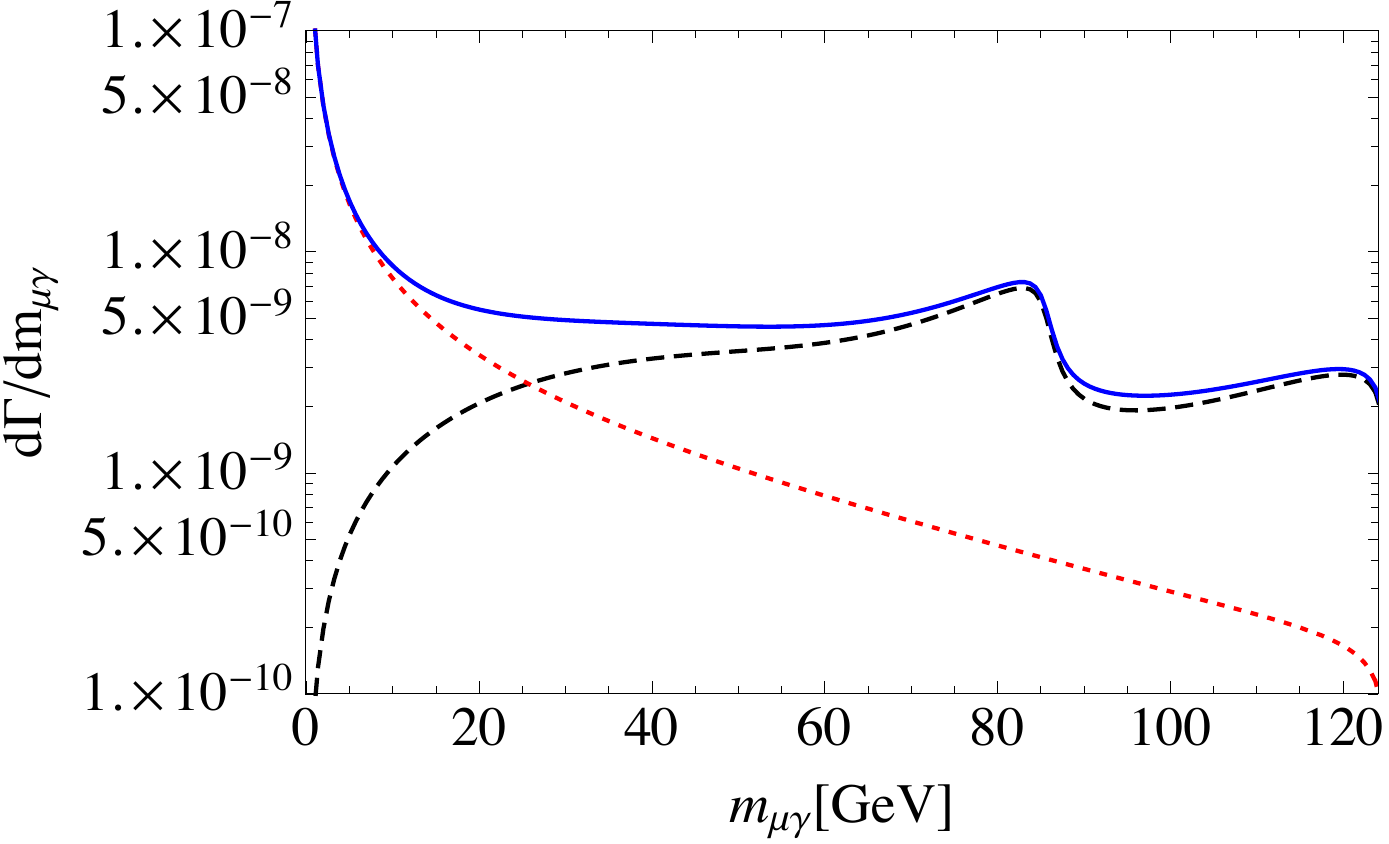}}
         \end{center}
         \caption{Differential decay rate with respect to the invariant mass
           $\sqrt{t}\equiv m_{\mu\gamma}$ of the muon-photon pair.  The
           tree-level, one-loop, and total contribution are denoted by
           red dotted, black dashed lines, and solid blue lines,
           respectively.  No cuts on $s$ are introduced.  }
	\label{Fig:4}
~\\[-3mm]\hrule
\end{figure}

The numerical values of the rates are
\begin{align}
\Gamma^{(e)}= 0.237\,\text{keV}\,,\qquad \Gamma^{(\mu)}= 0.262\,\text{keV}. 
\label{gemu}
\end{align}
The difference between $\Gamma^{(e)}$ and $ \Gamma^{(\mu)}$ stems from
the tree-level contribution. \fig{Fig:4} shows the differential decay
rate with respect to the invariant mass of the lepton-photon pair. 
With a total Higgs width of $4.1\mev$ the rates in \eq{gemu} correspond
to the branching ratios
\begin{align}
  B(H\to e \bar e \gamma)=5.8\cdot 10^{-5}, \qquad
  B(H\to \mu \bar \mu  \gamma)=6.4\cdot 10^{-5}. 
\label{bremu}
\end{align}
These branching ratios are roughly three times smaller than
$ B(H\to \mu \bar \mu) $.

\subsection{Forward-backward asymmetry}
\begin{figure}[tb]
\hrule
~\\[2mm]
	\begin{center}
		\subfigure{\includegraphics[width=0.52\textwidth]{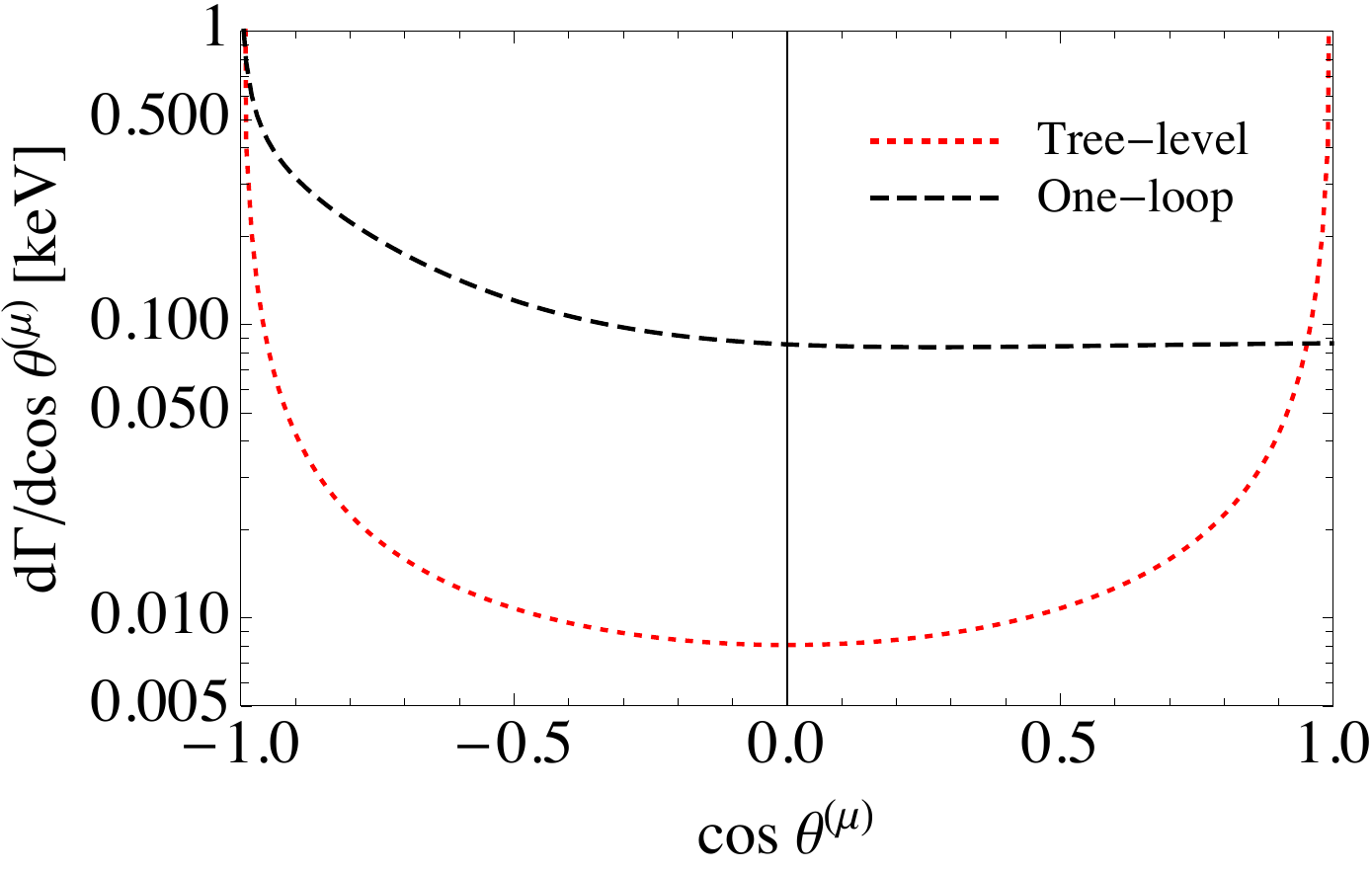}}
         \end{center}
         \caption{Differential decay rate with respect to
           $\cos\theta^{(\mu)}$, where $\theta^{(\mu)}$ is the angle
           between the lepton and the photon in the rest frame of the
           Higgs boson. For the integration over $m_{\mu\mu}$ we apply
           the cuts $m_{\mu\mu} > 0.1 m_H$ and
           $E_{\gamma, \text{min}} = 5\,\text{GeV}$.  }
\label{Fig:FB-distribution}
~\\[-3mm]\hrule
\end{figure}

Here we present the differential decay distribution with
respect to $\cos\theta^{(\ell)}$, where $\theta^{(\ell)}$ is the angle
between lepton and the photon in the rest frame of the Higgs boson,
$t=E_\gamma(E_1 -\vert\vec{p}_1\vert \cos\theta^{(\ell)})$. The
resulting distribution for the case of $\ell=\mu$ is shown in 
\fig{Fig:FB-distribution}. For this evaluation we apply the cuts
$m_{\mu\mu} > 0.1 m_H$ and $E_{\gamma} > 5\,\text{GeV}$ (and no cuts on
$E_{1,2}$).
We define the forward-backward asymmetry  with respect to $\theta^{(\ell)}$ as
\begin{align}
  \mathcal{A}^{(\ell)}_{\text{FB}}=\frac{\int_{-1}^0
  \frac{d\Gamma}{d\cos\theta^{(\ell)}} - \int_{0}^1 \frac{d\Gamma}{d\cos\theta^{(\ell)}}}{ 
  \int_{-1}^0 \frac{d\Gamma}{d\cos\theta^{(\ell)}} + 
  \int_{0}^1 \frac{d\Gamma}{d\cos\theta^{(\ell)}}}\,.\label{AFB}
\end{align}
With the cuts $m_{\ell\ell} > 0.1 m_H$ and
$E_{\gamma, \text{min}} = 5\,\text{GeV}$, applied to both the numerator
and the denominator in \eq{AFB}, we obtain the numerical values:
\begin{align}
  \mathcal{A}^{(e)}_{\text{FB}}=0.343\,,\qquad \mathcal{A}^{(\mu)}_{\text{FB}}=0.255\,.
  \label{afbres}
\end{align}

We note that the contribution of the tree-level diagrams to the
numerators of the asymmetries in \eq{AFB} can be neglected relatively to the
dominant one-loop contribution and the numerators for the electron and
muon case are essentially identical. Thus the 
difference between $ \mathcal{A}^{(e)}_{\text{FB}}$ and
$\mathcal{A}^{(\mu)}_{\text{FB}}$ in \eq{afbres} stems from the denominators; 
the non-negligible tree-level contribution increases the full rate
in the muon case. The difference between  $ \mathcal{A}^{(e)}_{\text{FB}}$ and
$\mathcal{A}^{(\mu)}_{\text{FB}}$ is numerically more pronounced than
the one between $\Gamma^{(e)}$ and $ \Gamma^{(\mu)}$ in \eq{gemu},
because different cuts are used in \eqsand{gemu}{afbres}.

\subsection{Comparison with previous results}
\begin{figure}[tb]
\hrule
	\begin{center}
		\subfigure{\includegraphics[width=0.7\textwidth]{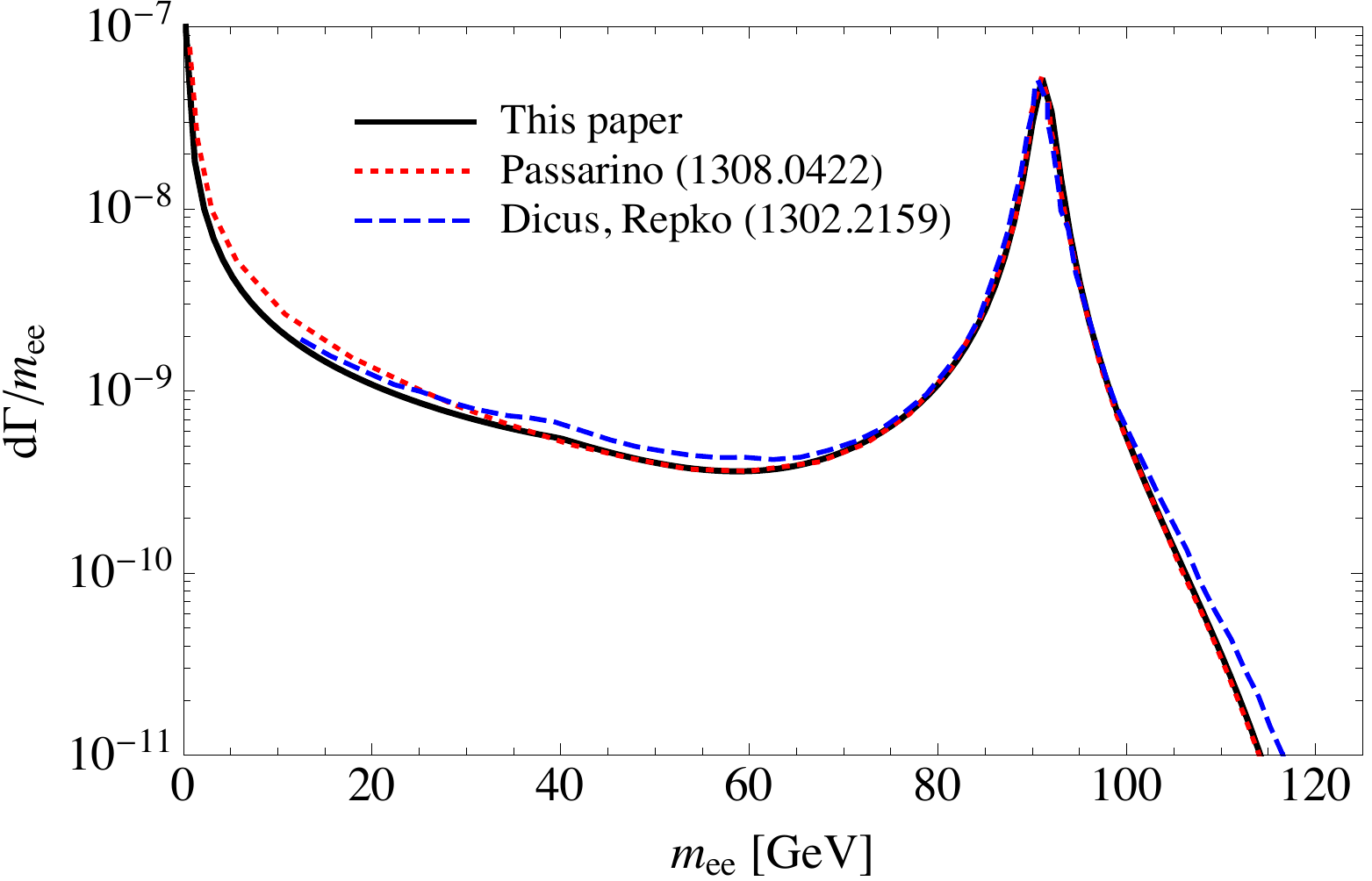}}
		\hspace{.1cm}
         \end{center}
         \caption{Differential decay rate with respect to the invariant dilepton mass
           for electrons. Our result is 
           denoted by a black solid line, while the results of
           Refs.~\cite{Passarino:2013nka} and \cite{Dicus:2013ycd} are 
           denoted by red short-dashed  and blue long-dashed lines, 
           respectively.}\label{Fig:7}
~\\[-3mm]\hrule
\end{figure}

The main goal of our paper is the resolution of the discrepancies
between the different results in the literature.  Only Abbasabadi et al.
\cite{Abbasabadi:1996ze} and Chen, Qiao and Zhu \cite{Chen:2012ju} provided an
analytic result. The 
latter paper is the only one containing terms with the Levi-Civita 
tensor and the origin and cancellation of such terms is discussed 
above in Sec.~\ref{amp}. In the case of the former paper we have 
numerically evaluated the presented formula (taking into account the 
typo reported in Ref.~\cite{Dicus:2013ycd}) and only find 
quantitative agreement in some regions, while we significantly disagree 
in others.  In the next step we have digitalized the plots 
for $d\Gamma (H\to \ell^+\ell^-\gamma)/d m_{ee}$ of
Refs.~\cite{Passarino:2013nka} and  \cite{Dicus:2013ycd}, which  
have used the same cuts on the kinematic variables. We compare the two
results and ours in \fig{Fig:7}. 
From this plot we can see that the difference between the previous works
in Refs.~\cite{Passarino:2013nka,Dicus:2013ycd} is up to 30\%. Our
result is close to the one of Ref.~\cite{Passarino:2013nka} for
$m_{ee}\gtrsim 40\gev$, but significantly deviates for smaller values of
$m_{ee}$.  One may speculate that the choice of the QED fine structure
constant $\alpha$, which fixes the $e^4$ term in the overall
normalization constant $\mathcal{N}$ in \eq{norm} might account for the
difference in at least the region with $m_{ee}\gtrsim 35\gev$. 
Shifting $\alpha^{-1}$ from the value in \eq{inp} to
  $\alpha^{-1}=128$ only alleviates the tension in \fig{Fig:7} for
  $m_{ee}\lesssim 40\gev $, but does
not fully resolve it. Furthermore, the agreement with the total
  decay rate becomes worse.
Furthermore the shape of the
distributions is different, and eventually the numerical integration
over $t$ is the reason for this discrepancy. We have cross-checked our
result by using different Monte-Carlo generators, i.e.\ \emph{Vegas}\
and the one implemented in \emph{Mathematica}. Our
  result for the integrated rate $\Gamma^{(e)}=0.237\,\text{keV}$ is in reasonable
  agreement with the result $\Gamma^{(e)}=0.233\,\text{keV}$ given in
  Ref. \cite{Passarino:2013nka}. We further remark that we disagree
with Ref.~\cite{Passarino:2013nka} in the tree-level contribution to the
integrated decay rate of $H\to \mu\bar\mu \gamma$ amplitude by a factor of 2.

\begin{figure}[tb]
\hrule
	\begin{center}
		\subfigure{\includegraphics[width=0.7\textwidth]{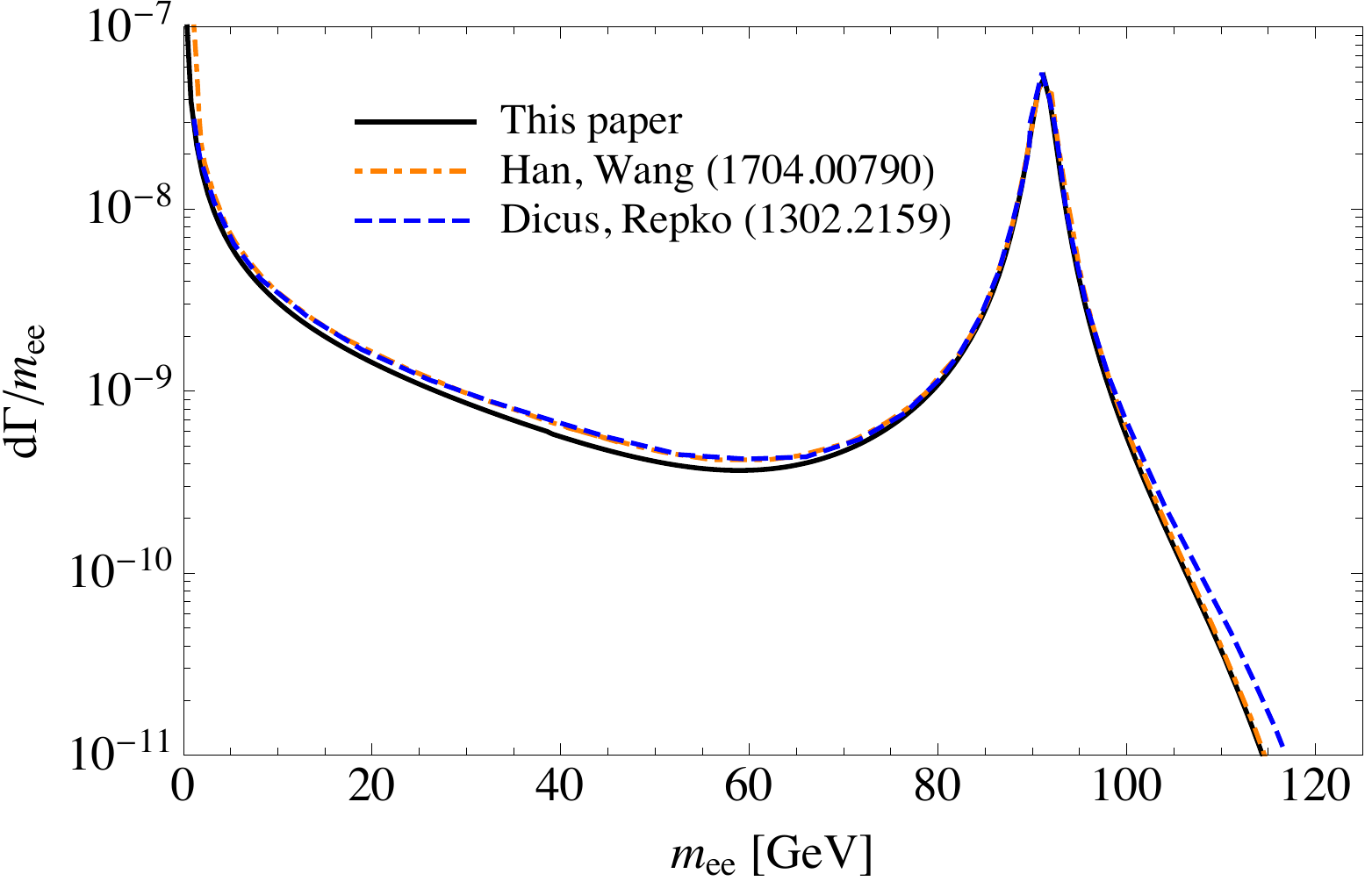}}
		\hspace{.1cm}
         \end{center}
         \caption{Differential decay rate with respect to the invariant dilepton mass
           for the electron case. Our result is 
           denoted by a black solid line, while the results of
           Refs.~\cite{Dicus:2013ycd} and \cite{Han:2017yhy} are 
           denoted by blue dashed and orange dash-dotted lines, 
           respectively.}\label{fig:comp2}
~\\[-3mm]\hrule
\end{figure}

Reference \cite{Han:2017yhy} further presents results for $d\Gamma (H\to \ell^+\ell^-\gamma)/d m_{ee}$ for a different choice of cuts, namely:
\begin{align}
\Delta R_{\gamma e^+} > 0.4\,,\qquad \Delta R_{\gamma e^-} >0.4\,,\label{cutsHanWang}
\end{align}
where $\Delta R_{\gamma f} = (\Delta\eta^2 + \Delta\phi^2)^{1/2}$
denotes the rapidity-azimuthal angle separation.  Digitalizing the plot
in this paper as well, we compare the presented results with ours in
\fig{fig:comp2}. We have found that the effect of the cuts in Eq.
\eqref{cutsHanWang} does not alter the loop-induced distribution by
more than $2\%$ in the region where we observe deviations from
Ref.~\cite{Han:2017yhy}. Therefore, we add in the same figure the digitalized
result of Ref. \cite{Dicus:2013ycd} that does not employ any cut. We
observe good agreement between Refs.~\cite{Dicus:2013ycd,Han:2017yhy}
{in a region} below the $Z$ peak, where we agree with these results
well for $m_{\mu\mu}\gtrsim 70 \gev$, while deviating
otherwise. 
For $m_{\mu\mu}> M_Z$ we agree well with the result of
Ref.~\cite{Han:2017yhy}.

\section{Conclusions\label{conc}}
The results in the literature for the differential decay rates
$d\Gamma (H\to \ell^+\ell^-\gamma)/d m_{\ell\ell}$ with $\ell=e,\mu$,
differ substantially. We have performed a new calculation of the
differential decay rate
$d^2\Gamma (H\to \ell^+\ell^-\gamma)/(d s\, dt)$, where $s$ is the
squared invariant mass of the lepton-antilepton pair and $t$ is the
corresponding quantity for the lepton-photon pair.   We have performed our
calculation in an $R_\xi$ gauge and have verified the gauge independence
of the result. After presenting various differential decay distributions
we have studied the forward-backward asymmetry defined in terms of the flight
direction of the photon with respect to the lepton. {These asymmetries,
quoted in \eq{afbres}, are sizable.}

{In experimental studies one defines cuts in the laboratory reference
  frame rather than the Higgs rest frame and the comparison between data
  and SM prediction requires the use of the fully differential decay
  rate. For this purpose we present an analytic expression  in a
compact form in Appendix~\ref{sec:res} and provide ancillary files
for the use by experimental collaborations.}

\section*{Acknowledgements}
We thank Vladyslav Shtabovenko for useful discussions and valuable
  insights into
\emph{FeynCalc} as well as Wayne Repko, Giampero Passarino, and Xing Wang
for helpful correspondence on
Refs.~\cite{Abbasabadi:1996ze,Dicus:2013ycd,Passarino:2013nka,Han:2017yhy}.
The research of U.N.\ is supported by BMBF under grant no.~05H2018
(ErUM-FSP T09) - \emph{BELLE II: Theoretische Studien zur
  Flavourphysik}. A.K.\ acknowledges the support from the doctoral
school \emph{KSETA}\ and the \emph{Graduate School Scholarship
  Programme}\ of the \emph{German Academic Exchange Service (DAAD)}.

\begin{appendix}
\section{Explicit results}\label{sec:res}
In this section we quote the formulas for $a_1$ and $b_1$,
which characterize the decay rate in \eq{loop-decayrate} and are
  introduced in \eq{loop_amp},  in terms of
  the coefficient functions appearing in the Passarino-Veltman
  decomposition of the tensor integrals. We follow the conventions of
  \emph{FeynCalc}\ \cite{Shtabovenko:2016sxi, Mertig:1990an}.
In the equations below we use $D\equiv 4-2\epsilon$ which appears 
in the coefficients of the UV-divergent loop function $B_0$. The products
are understood to be expanded in $\epsilon$ to order $\epsilon^0$.
One easily verifies that the $1/\epsilon$ pole vanishes from $a_1$ and 
$b_1$ in the sum of the various terms involving $B_0$.

The function $a_1$ reads 
\begin{align}
a_1 &= \frac{e^4}{(2\pi)^{2}}\bigg[\frac{8 m_t^2\big((D-4)m_H^2-(D-2)s\big)B_0(m_H^2,m_t^2,m_t^2)}{3 m_W \sin\theta_W (D-2) s (m_H^2-s)^2}+\frac{16 m_t^2 B_0(s,m_t^2,m_t^2)}{3 m_W \sin\theta_W (D-2)(m_H^2-s)^2}\nonumber \\
&-\frac{\big(2(D-1)m_W^2+m_H^2\big)\big((D-4)m_H^2-(D-2)s\big)B_0(m_H^2,m_W^2,m_W^2)}{2\,m_W\sin\theta_W (D-2) s (m_H^2-s)^2}\nonumber\\
&-\frac{\big(2(D-1)m_W^2+m_H^2\big)B_0(s,m_W^2,m_W^2)}{m_W \sin\theta_W (D-2) (m_H^2-s)^2}\nonumber\\
&-\frac{2 m_t^2\big(-2m_H^2+ 2 s+8 m_t^2\big) C_0(0, m_H^2, s, m_t^2, m_t^2, m_t^2)}{3 \,m_W\sin\theta_W\, s (m_H^2-s)}\nonumber\\
&+\frac{m_W\big(-3 m_H^2+6 m_W^2+4 s\big)C_0(0,m_H^2,s,m_W^2,m_W^2,m_W^2)}{\sin\theta_W\,s(m_H^2-s)}\nonumber\\
&+\frac{m_Z\sin\theta_W}{\cos^3\theta_W} \big(D_{23}(0, u, 0, t, 0, m_H^2, 0, m_Z^2, 0, m_Z^2)+(D_{23}\to D_{33})\big)\nonumber\\
&+\frac{1}{s-m_Z^2 + i m_Z \Gamma_Z}\bigg(\frac{(4\cos\theta^2_W-1)m_Z C_0(0,m_H^2,s,m_W^2,m_W^2,m_W^2)}{\cos\theta_W \sin\theta_W}\nonumber\\
& + \frac{(5-8\cos\theta_W^2)m_t^2 C_0(0,m_H^2,s,m_t^2,m_t^2,m_t^2)}{6\cos^2\theta_W \sin\theta_W m_W} + \frac{2(8\cos\theta^2_W-5)m_t^2 C_{12}(0, m_H^2, s, m_t^2, m_t^2, m_t^2)}{3\cos^2\theta_W \sin\theta_W m_W} \nonumber\\
&-\frac{\big(2m_W^2(6\cos^2\theta_W-1)+(2\cos^2\theta_W-1)m_H^2\big)C_{12}(0,m_H^2,s,m_W^2,m_W^2,m_W^2)}{2\,\cos^2\theta_W\sin\theta_W m_W} \bigg)\bigg]\label{a1}
\end{align}
The explicit form of the function $b_1$ is:
\begin{align}
b_1 &= \frac{e^4}{(2\pi)^{2}}\bigg[\frac{8 \big((D-4)m_H^2-(D-2)s\big)B_0(m_H^2,m_t^2,m_t^2)m_t^2}{3\, m_W \sin\theta_W (D-2)(m_H^2-s)^2s}+\frac{16 B_0(s,m_t^2,m_t^2)m_t^2}{3 m_W\sin\theta_W (D-2)(m_H^2-s)^2}\nonumber\\
&-\frac{(m_H^2+2(D-1)m_W^2) B_0(s,m_W^2,m_W^2)}{m_W\sin\theta_W (D-2)(m_H^2-s)^2}\nonumber\\
&-\frac{\big(m_H^2+2(D-1)m_W^2\big) \big((D-4)m_H^2-(D-2)s\big)B_0(m_H^2,m_W^2,m_W^2)}{2\,m_W\sin\theta_W(D-2) (m_H^2-s)^2 s}\nonumber\\
&+\frac{2\big(2 m_H^2-8m_t^2- 2\,s\big)C_0(0,m_H^2, s, m_t^2, m_t^2, m_t^2)m_t^2}{3 m_W \sin\theta_W (m_H^2-s)s}\nonumber\\
&+\frac{m_W \big(-3 m_H^2+6 m_W^2+ 4\,s\big) C_0(0,m_H^2, s, m_W^2, m_W^2, m_W^2)}{\sin\theta_W\,(m_H^2-s)s}\nonumber\\
&- \frac{m_W D_0(m_H^2,0,0,0,s,t,m_W^2,m_W^2,m_W^2,0)}{2\sin^3\theta_W} - \frac{m_W D_1(0, t, m_H^2, s,0,0, m_W^2,0, m_W^2, m_W^2)}{2\,\sin^3\theta_W}\nonumber\\
&+ \frac{m_W D_3(0, u, m_H^2, s, 0, 0 ,m_W^2,0,m_W^2,m_W^2)}{2\,\sin^3\theta_W}
-\frac{m_W D_3(0,t, m_H^2,s, 0, 0, m_W^2,0,m_W^2,m_W^2)}{2\,\sin^3\theta_W}\nonumber\\
&+\frac{m_W D_{12}(0, t, m_H^2, s, 0, 0, m_W^2,0,m_W^2,m_W^2)}{2\,\sin^3\theta_W}\nonumber\\
&+\frac{m_Z (1-2\cos^2\theta_W)^2D_{23}(0,u,0,t,0,m_H^2,0,m_Z^2,0,m_Z^2)}{4\,\cos^3\theta_W \sin^3\theta_W}\nonumber\\
&+ \frac{m_W D_{23}(0,u,m_H^2,s,0,0,m_W^2,0,m_W^2,m_W^2)}{2\sin^3\theta_W}\nonumber\\
&+ \frac{m_W D_{23}(0,t,m_H^2,s,0,0,m_W^2,0,m_W^2,m_W^2)}{2\sin^3\theta_W}\nonumber\\
&+\frac{m_Z (1-2\cos^2\theta_W)^2 D_{33}(0, u,0,t,0,m_H^2,0,m_Z^2,0,m_Z^2)}{4\cos^3\theta_W \sin^3\theta_W}\nonumber\\
&-\frac{1}{s-m_Z^2+i m_Z \Gamma_Z}\bigg(\frac{(2\cos^2\theta_W-1)(4\cos^2\theta_W-1)m_Z C_0(0, m_H^2,s,m_W^2,m_W^2,m_W^2)}{2\cos\theta_W \sin^3\theta_W}\nonumber\\
&+\frac{(5-8\cos^2\theta_W)(2\cos^2\theta_W-1)m_t^2 C_0(0, m_H^2,s, m_t^2, m_t^2, m_t^2)}{12\,\cos^2\theta_W\sin^3\theta_W m_W}\nonumber\\
&+ \frac{(2\cos^2\theta_W-1)(8\cos^2\theta_W-5)m_t^2 C_{12}(0, m_H^2, s, m_t^2, m_t^2, m_t^2)}{3\cos^2\theta_W\sin^3\theta_W m_W} \nonumber\\
&- \frac{(2\cos^2\theta_W-1)\big(2m_W^2(6\cos^2\theta_W-1)+(2\cos^2\theta_W-1)m_H^2\big)C_{12}(0, m_H^2, s, m_W^2, m_W^2, m_W^2)}{4\cos^2\theta_W\sin^3\theta_W m_W}\bigg)\bigg].\label{b1}
\end{align}

\section{Usage of the ancillary files}

In the ancillary files attached to the arXiv preprint we provide the
analytical expressions for the one-loop coefficients $a_i, b_i$,
$i=1,2$, defined in Eq. \eqref{loop_amp}. The file
\verb|coeff_nonreduced.m| contains the results that match Eqs.
\eqref{a1}, \eqref{b1}, while \verb|coeff_reduced.m| contains the same
results in the form that is reduced to the standard basis
$\{A_0,B_0,C_0,D_0 \}$ of scalar one-loop functions. The coefficients
are provided in both \emph{FeynCalc} and |\emph{PackageX} notations and
conventions. After importing the files, the coefficients can be called
using, for example, \verb|a1PXr| for the coefficient $a_1$ in the
reduced form and \emph{PackageX} notation, or \verb|b1FCnr| for coefficient
$b_1$ in the nonreduced form and \emph{FeynCalc} notation. For instance, the
coefficients given in the \emph{Package-X} notation can be numerically
evaluated using \emph{Collier} \cite{Denner:2016kdg} via the \emph{CollierLink}
\cite{Patel:2015tea} package; the detailed manual for the latter package
can be found in \cite{CollierLink}.
\end{appendix}


\begin{thebibliography}{9}
\bibitem{Chatrchyan:2012xdj}
  S.~Chatrchyan {\it et al.} [CMS Collaboration],
  Phys.\ Lett.\ B {\bf 716} (2012) 30
  [arXiv:1207.7235 [hep-ex]].

\bibitem{Aad:2012tfa}
  G.~Aad {\it et al.} [ATLAS Collaboration],
  Phys.\ Lett.\ B {\bf 716} (2012) 1
  [arXiv:1207.7214 [hep-ex]].

\bibitem{Englert:2014uua}
  C.~Englert, A.~Freitas, M.~M.~Mühlleitner, T.~Plehn, M.~Rauch, M.~Spira and K.~Walz,
  J.\ Phys.\ G {\bf 41} (2014) 113001
  [arXiv:1403.7191 [hep-ph]].

\bibitem{Aad:2015ona}
  G.~Aad {\it et al.} [ATLAS Collaboration],
  JHEP {\bf 1508} (2015) 137
  [arXiv:1506.06641 [hep-ex]].
\bibitem{ATLAS:2014aga}
  G.~Aad {\it et al.} [ATLAS Collaboration],
  Phys.\ Rev.\ D {\bf 92} (2015) no.1,  012006
  [arXiv:1412.2641 [hep-ex]].
\bibitem{Sirunyan:2018egh}
  A.~M.~Sirunyan {\it et al.} [CMS Collaboration],
  Phys.\ Lett.\ B {\bf 791} (2019) 96
  [arXiv:1806.05246 [hep-ex]].
\bibitem{Sirunyan:2019twz}
  A.~M.~Sirunyan {\it et al.} [CMS Collaboration],
  Phys.\ Rev.\ D {\bf 99} (2019) no.11,  112003
  [arXiv:1901.00174 [hep-ex]].

\bibitem{Chatrchyan:2013mxa}
  S.~Chatrchyan {\it et al.} [CMS Collaboration],
  Phys.\ Rev.\ D {\bf 89} (2014) no.9,  092007
  [arXiv:1312.5353 [hep-ex]].
\bibitem{Aad:2014tca}
  G.~Aad {\it et al.} [ATLAS Collaboration],
  Phys.\ Lett.\ B {\bf 738} (2014) 234
  [arXiv:1408.3226 [hep-ex]].
\bibitem{Aaboud:2018ezd}
  M.~Aaboud {\it et al.} [ATLAS Collaboration],
  Phys.\ Lett.\ B {\bf 786} (2018) 114
  [arXiv:1805.10197 [hep-ex]].

\bibitem{Aad:2015vsa}
  G.~Aad {\it et al.} [ATLAS Collaboration],
  JHEP {\bf 1504} (2015) 117
  [arXiv:1501.04943 [hep-ex]].
  
\bibitem{Aaboud:2018pen}
  M.~Aaboud {\it et al.} [ATLAS Collaboration],
  Phys.\ Rev.\ D {\bf 99} (2019) 072001
  [arXiv:1811.08856 [hep-ex]].
\bibitem{Sirunyan:2017khh}
  A.~M.~Sirunyan {\it et al.} [CMS Collaboration],
  Phys.\ Lett.\ B {\bf 779} (2018) 283
  [arXiv:1708.00373 [hep-ex]].

\bibitem{Sirunyan:2018kst}
  A.~M.~Sirunyan {\it et al.} [CMS Collaboration],
  Phys.\ Rev.\ Lett.\  {\bf 121} (2018) no.12,  121801
  [arXiv:1808.08242 [hep-ex]].
  
\bibitem{Aaboud:2018zhk}
  M.~Aaboud {\it et al.} [ATLAS Collaboration],
  Phys.\ Lett.\ B {\bf 786} (2018) 59
  [arXiv:1808.08238 [hep-ex]].

\bibitem{Sirunyan:2018shy}
  A.~M.~Sirunyan {\it et al.} [CMS Collaboration],
  JHEP {\bf 1808} (2018) 066
  [arXiv:1803.05485 [hep-ex]].
  
\bibitem{Aaboud:2018urx}
  M.~Aaboud {\it et al.} [ATLAS Collaboration],
  Phys.\ Lett.\ B {\bf 784} (2018) 173
  [arXiv:1806.00425 [hep-ex]].

\bibitem{Sirunyan:2018ouh}
  A.~M.~Sirunyan {\it et al.} [CMS Collaboration],
  JHEP {\bf 1811} (2018) 185
  [arXiv:1804.02716 [hep-ex]].
\bibitem{Aaboud:2018xdt}
  M.~Aaboud {\it et al.} [ATLAS Collaboration],
  Phys.\ Rev.\ D {\bf 98} (2018) 052005
  [arXiv:1802.04146 [hep-ex]].

\bibitem{Eberhardt:2012gv}
  O.~Eberhardt, G.~Herbert, H.~Lacker, A.~Lenz, A.~Menzel, U.~Nierste and M.~Wiebusch,
  Phys.\ Rev.\ Lett.\  {\bf 109} (2012) 241802
  [arXiv:1209.1101 [hep-ph]].

\bibitem{Eberhardt:2013uba}
  O.~Eberhardt, U.~Nierste and M.~Wiebusch,
  JHEP {\bf 1307} (2013) 118
  [arXiv:1305.1649 [hep-ph]].

\bibitem{Belanger:2013xza}
  G.~Belanger, B.~Dumont, U.~Ellwanger, J.~F.~Gunion and S.~Kraml,
  Phys.\ Rev.\ D {\bf 88} (2013) 075008
  [arXiv:1306.2941 [hep-ph]].

\bibitem{Baglio:2014nea}
  J.~Baglio, O.~Eberhardt, U.~Nierste and M.~Wiebusch,
  Phys.\ Rev.\ D {\bf 90} (2014) no.1,  015008
  [arXiv:1403.1264 [hep-ph]].

\bibitem{Kanemura:2014bqa}
  S.~Kanemura, K.~Tsumura, K.~Yagyu and H.~Yokoya,
  Phys.\ Rev.\ D {\bf 90} (2014) 075001
  [arXiv:1406.3294 [hep-ph]].

\bibitem{Dev:2014yca}
  P.~S.~Bhupal Dev and A.~Pilaftsis,
  JHEP {\bf 1412} (2014) 024
   Erratum: [JHEP {\bf 1511} (2015) 147]
  [arXiv:1408.3405 [hep-ph]].

\bibitem{Broggio:2014mna}
  A.~Broggio, E.~J.~Chun, M.~Passera, K.~M.~Patel and S.~K.~Vempati,
  JHEP {\bf 1411} (2014) 058
  [arXiv:1409.3199 [hep-ph]].

\bibitem{Chowdhury:2015yja}
  D.~Chowdhury and O.~Eberhardt,
  JHEP {\bf 1511} (2015) 052
  [arXiv:1503.08216 [hep-ph]].

\bibitem{Bernon:2015qea}
  J.~Bernon, J.~F.~Gunion, H.~E.~Haber, Y.~Jiang and S.~Kraml,
  Phys.\ Rev.\ D {\bf 92} (2015) no.7,  075004
  [arXiv:1507.00933 [hep-ph]].

\bibitem{Haber:2015pua}
  H.~E.~Haber and O.~Stål,
  Eur.\ Phys.\ J.\ C {\bf 75} (2015) no.10,  491
   Erratum: [Eur.\ Phys.\ J.\ C {\bf 76} (2016) no.6,  312]
  [arXiv:1507.04281 [hep-ph]].

\bibitem{Han:2017pfo}
  L.~Wang, F.~Zhang and X.~F.~Han,
  Phys.\ Rev.\ D {\bf 95} (2017) no.11,  115014
  [arXiv:1701.02678 [hep-ph]].

\bibitem{Chowdhury:2017aav}
  D.~Chowdhury and O.~Eberhardt,
  JHEP {\bf 1805} (2018) 161
  [arXiv:1711.02095 [hep-ph]].



\bibitem{Keshavarzi:2018mgv}
  A.~Keshavarzi, D.~Nomura and T.~Teubner,
  Phys.\ Rev.\ D {\bf 97} (2018) no.11,  114025
  [arXiv:1802.02995 [hep-ph]].
  
  
\bibitem{Aaij:2019wad}   
  R.~Aaij {\it et al.} [LHCb Collaboration],
  Phys.\ Rev.\ Lett.\  {\bf 122} (2019) no.19,  191801
  [arXiv:1903.09252 [hep-ex]].
  
\bibitem{Aaij:2017vbb}
  R.~Aaij {\it et al.} [LHCb Collaboration],
  JHEP {\bf 1708} (2017) 055
  [arXiv:1705.05802 [hep-ex]].

\bibitem{Alguero:2019ptt}
  M.~Algueró, B.~Capdevila, A.~Crivellin, S.~Descotes-Genon, P.~Masjuan, J.~Matias and J.~Virto,
  Eur.\ Phys.\ J.\ C {\bf 79} (2019) no.8,  714
  [arXiv:1903.09578 [hep-ph]].

\bibitem{Aebischer:2019mlg}
  J.~Aebischer, W.~Altmannshofer, D.~Guadagnoli, M.~Reboud, P.~Stangl and D.~M.~Straub,
  arXiv:1903.10434 [hep-ph].
  
\bibitem{Sirunyan:2018tbk}
  A.~M.~Sirunyan {\it et al.} [CMS Collaboration],
  JHEP {\bf 1811} (2018) 152
  [arXiv:1806.05996 [hep-ex]].

\bibitem{Abbasabadi:1996ze} 
  A.~Abbasabadi, D.~Bowser-Chao, D.~A.~Dicus and W.~W.~Repko,
  Phys.\ Rev.\ D {\bf 55}, 5647 (1997)
  [hep-ph/9611209].
  
\bibitem{Chen:2012ju}
  L.~B.~Chen, C.~F.~Qiao and R.~L.~Zhu,
  Phys.\ Lett.\ B {\bf 726} (2013) 306
  [arXiv:1211.6058 [hep-ph]].

\bibitem{Dicus:2013ycd}
  D.~A.~Dicus and W.~W.~Repko,
  Phys.\ Rev.\ D {\bf 87} (2013) no.7,  077301
  [arXiv:1302.2159 [hep-ph]].
  
\bibitem{deFlorian:2016spz}
  D.~de Florian {\it et al.} [LHC Higgs Cross Section Working Group],
  arXiv:1610.07922 [hep-ph].
  
\bibitem{Passarino:2013nka}
  G.~Passarino,
  Phys.\ Lett.\ B {\bf 727} (2013) 424
  [arXiv:1308.0422 [hep-ph]].

\bibitem{Han:2017yhy}
  T.~Han and X.~Wang,
  JHEP {\bf 1710} (2017) 036
  [arXiv:1704.00790 [hep-ph]].
  
\bibitem{Denner:1999gp}
  A.~Denner, S.~Dittmaier, M.~Roth and D.~Wackeroth,
  Nucl.\ Phys.\ B {\bf 560} (1999) 33
  [hep-ph/9904472].
  
\bibitem{Denner:2006ic}
  A.~Denner and S.~Dittmaier,
  Nucl.\ Phys.\ Proc.\ Suppl.\  {\bf 160} (2006) 22
  [hep-ph/0605312].
  
\bibitem{Aaboud:2017uhw}
  M.~Aaboud {\it et al.} [ATLAS Collaboration],
  JHEP {\bf 1710} (2017) 112
  [arXiv:1708.00212 [hep-ex]].
\bibitem{Sirunyan:2017hsb}
  A.~M.~Sirunyan {\it et al.} [CMS Collaboration],
  JHEP {\bf 1809} (2018) 148
  [arXiv:1712.03143 [hep-ex]].
  
\bibitem{Brown:1952eu}
  L.~M.~Brown and R.~P.~Feynman,
  Phys.\ Rev.\  {\bf 85} (1952) 231.

\bibitem{Passarino:1978jh}
  G.~Passarino and M.~J.~G.~Veltman,
  Nucl.\ Phys.\ B {\bf 160} (1979) 151.

\bibitem{tHooft:1978jhc}
  G.~'t Hooft and M.~J.~G.~Veltman,
  Nucl.\ Phys.\ B {\bf 153} (1979) 365.

\bibitem{Denner:1991qq}
  A.~Denner, U.~Nierste and R.~Scharf,
  Nucl.\ Phys.\ B {\bf 367} (1991) 637.
  
\bibitem{Sun:2013rqa}
  Y.~Sun, H.~R.~Chang and D.~N.~Gao,
  JHEP {\bf 1305} (2013) 061
  [arXiv:1303.2230 [hep-ph]].
  
\bibitem{Abbasabadi:1995rc}
  A.~Abbasabadi, D.~Bowser-Chao, D.~A.~Dicus and W.~W.~Repko,
  Phys.\ Rev.\ D {\bf 52} (1995) 3919
  [hep-ph/9507463].
  
\bibitem{Hahn:2000kx}
  T.~Hahn,
  Comput.\ Phys.\ Commun.\  {\bf 140} (2001) 418
  [hep-ph/0012260].
  
\bibitem{Shtabovenko:2016sxi}
  V.~Shtabovenko, R.~Mertig and F.~Orellana,
  Comput.\ Phys.\ Commun.\  {\bf 207} (2016) 432
  [arXiv:1601.01167 [hep-ph]]
 
\bibitem{Mertig:1990an}
  R.~Mertig, M.~Bohm and A.~Denner,
  Comput.\ Phys.\ Commun.\  {\bf 64} (1991) 345.
  
\bibitem{Patel:2015tea}
  H.~H.~Patel,
  Comput.\ Phys.\ Commun.\  {\bf 197} (2015) 276
  [arXiv:1503.01469 [hep-ph]].
  
\bibitem{Shtabovenko:2016whf}
  V.~Shtabovenko,
  Comput.\ Phys.\ Commun.\  {\bf 218} (2017) 48
  [arXiv:1611.06793 [physics.comp-ph]].
  
  
\bibitem{CollierLink}
  H.~Patel, 
https://packagex.hepforge.org/Documentation/HTML/X/tutorial/LinkingToCOLLIER.html
    
\bibitem{Denner:2016kdg}
  A.~Denner, S.~Dittmaier and L.~Hofer,
  Comput.\ Phys.\ Commun.\  {\bf 212} (2017) 220
  [arXiv:1604.06792 [hep-ph]].
  
\bibitem{Denner:2005nn}
  A.~Denner and S.~Dittmaier,
  Nucl.\ Phys.\ B {\bf 734} (2006) 62
  [hep-ph/0509141].
  
\bibitem{Denner:2010tr}
  A.~Denner and S.~Dittmaier,
  Nucl.\ Phys.\ B {\bf 844} (2011) 199
  [arXiv:1005.2076 [hep-ph]].
  
\bibitem{Hahn:1998yk}
  T.~Hahn and M.~Perez-Victoria,
  Comput.\ Phys.\ Commun.\  {\bf 118} (1999) 153
 
\bibitem{LoopManual}
  T.~Hahn, http://www.feynarts.de/looptools/LT215Guide.pdf 
  
\bibitem{Lepage:1977sw}
  G.~P.~Lepage,
  J.\ Comput.\ Phys.\  {\bf 27} (1978) 192.
  
\bibitem{Hahn:2004fe}
  T.~Hahn,
  Comput.\ Phys.\ Commun.\  {\bf 168} (2005) 78
  [hep-ph/0404043].

\bibitem{Mathematica}
\textsc{Mathematica}, 
Wolfram Research Inc.,
Champaign, IL, 2019.
  
  
\end{thebibliography}
\end{document}